\RequirePackage{fix-cm}
\documentclass[twocolumn,epjc3]{svjour3}  % EPJC

\smartqed  % flush right qed marks, e.g. at end of proof
\RequirePackage[T1]{fontenc}
\usepackage{float,lscape}
\usepackage{flushend}
\usepackage{cuted}
\usepackage{graphicx}  % needed for figures
\usepackage{dcolumn}   % needed for some tables
\usepackage{bm}        % for math
\usepackage{amssymb}   % for math
\usepackage{sidecap}   % sidecaptions
\usepackage{subfigure}
\usepackage{amsmath}
\usepackage{pdfpages} % for authorlist
\usepackage{multirow}
\usepackage{xspace}
\usepackage{booktabs}
\usepackage{color}
\usepackage[normalem]{ulem}
\usepackage[numbers,sort&compress]{natbib}
\usepackage{epstopdf} %allows to use pdftex
\usepackage{placeins}

\usepackage{mathrsfs}
\usepackage{epsfig}
\usepackage{dcolumn}% Align table columns on decimal point
\usepackage{bm}% bold math
\usepackage{tabularx}
\usepackage[caption=false]{subfig}
\usepackage{breakurl}
\usepackage{wrapfig}
\usepackage{booktabs}
\usepackage{color}
\usepackage{slashed}
\usepackage{enumerate}
\usepackage{helvet}
\usepackage{bigstrut}
\usepackage{array}
\usepackage{rotating}
\usepackage{pdflscape}
\usepackage{afterpage}
\usepackage{enumerate}
\journalname{Eur. Phys. J. C}
\begin{document}
\title{\boldmath Measuring anomalous $Wtb$ couplings at $e^-p$ collider}
\author{Sukanta Dutta\thanksref{e1,addr1} \and Ashok Goyal\thanksref{e2,addr2}
\and Mukesh Kumar\thanksref{e3,addr3} \and Bruce Mellado\thanksref{e4,addr4}}  
\thankstext{e1}{e-mail: sukanta.dutta@gmail.com}
\thankstext{e2}{e-mail: agoyal45@yahoo.com}
\thankstext{e3}{e-mail: mukesh.kumar@cern.ch}
\thankstext{e4}{e-mail: bmellado@mail.cern.ch}

\institute{\emph{SGTB Khalsa College, University of
Delhi. Delhi-110007. India.}\label{addr1}
          \and
          \emph{Department of Physics and Astrophysics, University of
Delhi. Delhi-110007. India.}\label{addr2}
          \and
\emph{National Institute for Theoretical Physics, School of Physics, School of Physics and Mandelstam Institute for Theoretical Physics, University of the Witwatersrand, Johannesburg.}\label{addr3}
\and
\emph{University of the Witwatersrand, Private Bag 3, Wits 2050, Johannesburg, South Africa.} \label{addr4}
}
\date{}

\maketitle
\begin{abstract}
We study the accuracy with which the lowest order $CP$ conserving anomalous  $Wtb$ couplings in the single top quark production at the proposed large hadron electron collider (LHeC) can be probed. The one dimensional distribution of various kinematic observables at the parton level MC and their asymmetries arising due to the presence of anomalous couplings both in the hadronic and leptonic $W$ decay is examined.
\par We find that at 95 \% C.L. the anomalous coupling associated with the left handed vector current can be measured at an accuracy of the order of $\sim 10^{-2}-10^{-3}$, while those associated with the right handed vector and left as well as right handed  tensor  currents have sensitivity  at the order of  $\sim 10^{-1}-10^{-2}$  for the systematic uncertainty varying between 10\%-1\% at an integrated luminosity of 100 fb$^{-1}$. A comprehensive  analysis of  the combined covariance matrix derived from all one dimensional distributions of kinematical observables is used to compute the  errors in anomalous couplings.
\end{abstract}
%\pacs{}
 
\maketitle
\section{Introduction}
\label{sec:intro}
The top  quark  provides an excellent opportunity for the study of electroweak symmetry breaking mechanism as well as to provide glimpse of new physics (NP) beyond the standard model (SM). The top  quark decays almost exclusively in the $t\to b W^+$ channel. 

\par The  kinematic distributions of its  decayed particles from top  quark  provide the information about the $Wtb$ vertex and associated new physics potentiality with the top   quark production mechanism.
\begin{wrapfigure}{r}{0.3\textwidth}
  \centering
  \includegraphics[width=0.25\textwidth]{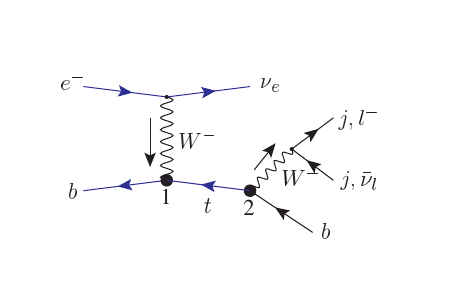}
  \caption{\small \em{Single anti-top quark production through charge current at the e-p collider.The blobs at vertices 1 and 2 show the effective  $W^-\bar t\bar b$ couplings, which includes the SM  contribution. Further $W^-$ decays into hadronic mode $via$ light quarks ($j\equiv \bar u,d,\bar c, s$) or leptonic mode ($l^-\equiv e^-,\mu^-$) with missing energy.}}
\label{fig:stop}
\end{wrapfigure}

\par Within the SM, the $Wtb$ vertex is purely left-handed, and its amplitude is given by the Cabibbo-Kobayashi-Maskawa (CKM) matrix element $V_{tb}$, related to weak interaction between a top and a $b$-quark and assuming $\left| V_{td} \right|^2 + \left| V_{ts} \right|^2 \ll \left| V_{tb} \right|^2$. The most general, lowest dimension, $CP$ conserving, Lagrangian for the $Wtb$ vertex is given by ~\cite{Kane:1991bg,  AguilarSaavedra:2008zc, AguilarSaavedra:2009mx}
%\begin{onecolumn}
\begin{eqnarray}
\mathscr L_{Wtb}& =& \frac{g}{\sqrt 2} \Big[ W_{\mu} \bar t \gamma^{\mu}( V_{tb}\, f_{1}^{L}  P_L + f_{1}^{R} P_R) b\nonumber\\ 
 &&- \frac{1}{2\, m_W} W_{\mu\nu} \bar t \sigma^{\mu\nu} ( f_{2}^{L} P_L + f_{2}^{R} P_R) b \Big] + h.c.
\nonumber\\\label{lagwtb}
\end{eqnarray}
%\end{onecolumn}
where $f_1^L\equiv 1+\Delta f_1^L$, $W_{\mu\nu} = \partial_{\mu} W_{\nu} - \partial_{\nu} W_{\mu}, P_{L,R} = \tfrac{1}{2} \left(1\mp \gamma_5 \right)$ are left- and right-handed projection operators, $\sigma^{\mu\nu}=i/2\left(\gamma^{\mu}\gamma^{\nu}-\gamma^{\nu}\gamma^{\mu}\right)$ and  $g = e/\sin \theta_{W}$. In SM  $\left\vert V_{tb} \right\vert \,f_1^L \simeq 1$, all  other couplings  $f_2^L, f_1^R, f_2^R$ vanish at tree level. Their non-vanishing values are generated at the one loop level \cite{Do:2002ky}.
\begin{figure}%{l}{0.25\textwidth}
%\begin{figure*}[h!]
  \centering
  \includegraphics[width=0.5\textwidth,clip]{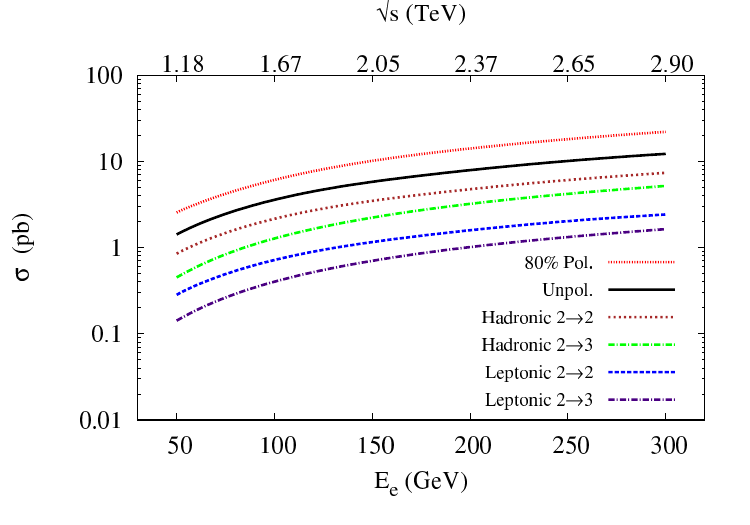} 
  \caption{\small \em{Single anti-top quark production cross section at the LHeC with the variation of electron energy $E_e$ and  fixed proton energy  $E_p$ = 7 TeV. The top two curves depict the cross-section for $e^-p\to \nu_e\, \bar t$ from the  80 \% polarized and  unpolarized $e^-$ beam, respectively.
The third and the fifth curve corresponds to the branching of the unpolarized cross-section into hadronic and leptonic decay modes of $W^-$. The first and the third curve from the below corresponds to the cross-section for $e^-p\to \bar t \,\nu_e\, b$ branching to the leptonic and hadronic decay modes of $W^-$, respectively. }}
\label{xsecvar}
\end{figure}
$Wtb$ anomalous couplings $f_i$ are constrained from flavor physics. 
The magnitudes of the right-handed vector and tensor couplings can be indirectly constrained from the measured branching ratio of the $b\to s \gamma$ process. Current 95$\%$ C.L. bounds based on the CLEO data give $\left\vert f_1^R \right\vert \leq 4.0 \times 10^{-3}$ at the 2-$\sigma$ level ~\cite{Larios:1999au,Burdman:1999fw,Barberio:2007cr}.  The branching ratio (BR) BR$(b\to s  \gamma)$ is computed by neglecting $\left\vert f_i\right\vert^2$ terms in the matrix element squared and assuming only one anomalous coupling to be non-zero at a time. The upper  and lower limits for $\left\vert V_{tb} \right\vert \,f_1^L$, $f_1^R,\, f_2^L$ and $f_2^R$ obtained from the $B$ decays are $-0.13 \le \left\vert V_{tb} \right\vert \Delta f_1^L \le 0.03$, $-0.0007 \le f_1^R \le 0.0025$, $-0.0015 \le f_2^L\le  0.0004$  and $-0.15 \le f_2^R\le  0.57$ , respectively \cite{Grzadkowski:2008mf}. If more than one coupling are taken non-zero simultaneously, their magnitudes in principle are not bound by $b\to s \gamma$ alone and the limits can be very different. Combining the analysis on $B_{d,s}=\bar B_{d,s}$ mixing and $B\to X_sl^+l^-$, authors of reference \cite{Drobnak:2011aa} constrained $Wtb$ couplings  within an effective field theory framework.  
\begin{figure}%{l}{0.25\textwidth}
%\begin{figure*}[h!]
  \centering
  \includegraphics[width=0.5\textwidth,clip]{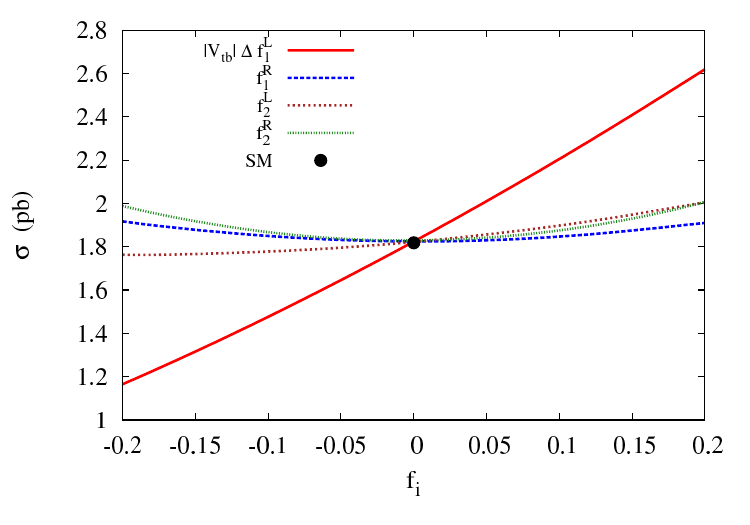} 
  \caption{\small \em{ Variation of the single anti-top quark production cross section with the effective $Wtb$ couplings (taking one anomalous coupling at a time with SM ) at the  production and decay  vertices, for fixed $E_p$ = 7 TeV and $E_e= 60$ GeV.  }}
\label{npxsecvar}
\end{figure}
\par The sensitivity of anomalous $Wtb$ couplings can also be measured from  $W^\pm$ helicity distributions arising from   top  decays to their  dominant $Wb$ mode in the top  pair production processes \cite{Abazov:2007ve}. It can also  be measured from  the observed single top  quark production cross section  through $W$-boson exchange and has both the linear and quadratic terms in the effective couplings. Although the single top production in the SM is comparable  to the $t\bar t$ pair production, it is quite challenging to make the extraction due to considerable backgrounds at the Tevatron \cite{Abazov:2009ii,CDF10793} and the LHC \cite{Chatrchyan:2012ep,Aad:2012ux}. D\O \,\,  with 5.4 fb$^{-1}$  data reported a combined analysis of   $W$ boson helicity studies and the single top quark production cross section  exclusively through $Wtb$ vertex. This sets upper limits on anomalous $Wtb$ couplings at 95 $\%$ C.L. {\it viz.}  $\left|f_2^L \right| \le 0.224$, $\left|f_1^R \right| \le 0.548$, $\left|f_2^R \right| \le 0.347$ (given in Table 1 of reference ~\cite{Abazov:2012uga}). Sensitivity of the anomalous $Wtb$ couplings on the cross-section of the associated $tW$ production are also studied at LHC through $\gamma p$ collision   at $\sqrt{s}$ =14 TeV for various luminosities and acceptance criterion \cite{Sahin:2012ry}.  The study of  coefficients of dimension six operators affecting $Wtb$ couplings from electroweak precision measurements \cite{Zhang:2012cd,Greiner:2011tt}, suggest that the upper limits on these couplings are one order of magnitude weaker, to those obtained directly from the helicity fraction study of  the top decay at NLO QCD \cite{Drobnak:2010ej}.
 \begin{figure*}[ht]
  \centering
  \begin{tabular}{ccc}
  \includegraphics[width=0.5\textwidth,clip]{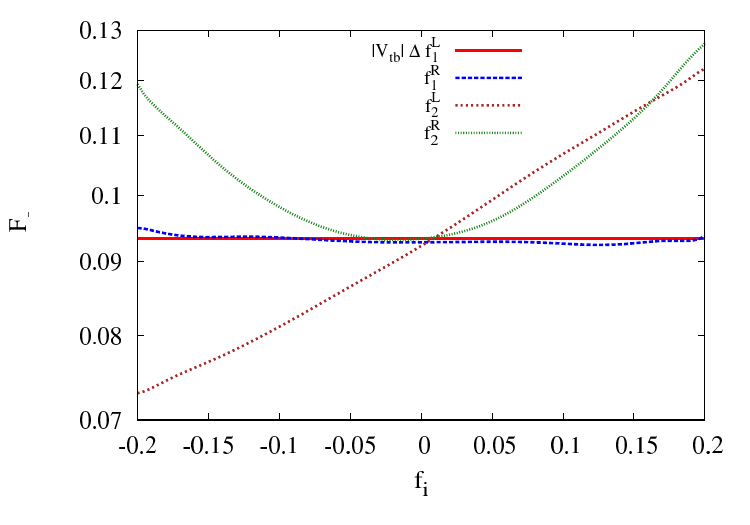} &
  \includegraphics[width=0.5\textwidth,clip]{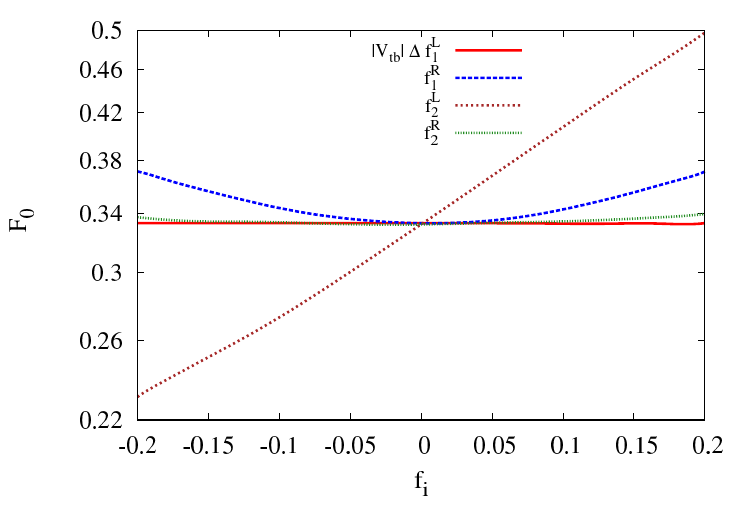}
  \end{tabular}
   \includegraphics[width=0.5\textwidth,clip]{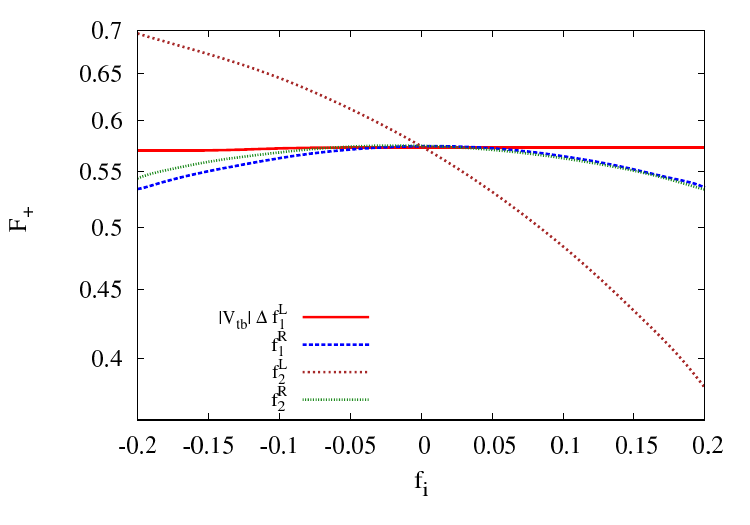}
  \caption{\small \em{The variation of helicity fractions ${\cal F}_-,\, {\cal F}_0$ and ${\cal F}_+$ as defined in the text with the anomalous coupling $f_i$.}}
\label{hel_ratios}
\end{figure*}

\par The sensitivity of the effective couplings in \eqref{lagwtb} can be studied through  one-dimensional distributions of  kinematic observables.
These distributions manifest a certain amount of associated asymmetry depending on the specific Lorentz structure, which can then be used as a discriminator to constraint  these anomalous couplings. Based on associated  asymmetries generated from the measured angular distributions   of $\cos \theta^* \footnotetext{The cosine of the angle $\theta^*$ between the momentum direction of the charged lepton from the W-boson decay and the reversed momentum direction of the b quark from top-quark decay, both boosted into the W-boson rest frame. }$  defined in ~\cite{AguilarSaavedra:2006fy}, the ATLAS collaboration ~\cite{ATLAS-37} set limits on single anomalous couplings at 95$\%$ C.L. to be $-0.44\le{\rm Re}\left(f_1^R \right) \le 0.48$, $-0.24\le {\rm Re}\left( f_2^L \right) \le 0.21$ and  $-0.49\le{\rm Re}\left( f_2^R \right) \le 0.15$. A combined constraint on  anomalous couplings from CMS and ATLAS \cite{atlasnote2013-33} shows the sensitivity of these couplings with respect to the helicity fraction  in the top quark decays. Constraints on $Wtb$ vertex based on the angular asymmetries constructed from ATLAS data and the t-channel single
top cross section in CMS have been analysed in \cite{AguilarSaavedra:2011ct}.
A projected sensitivity of all anomalous top couplings have also been studied in reference \cite{Bach:2012fb}. 
\par Effects of anomalous coupling on angular distributions of the $b$-quark and $\mu^+$ have been studied in $e^+ e^-$ linear collider with one specific semileptonic channel in the double resonance approximation for the $t$ and $\bar t$ production \cite{Cieckiewicz:2003sd, Grzadkowski:2002gt, Rindani:2000jg, Kolodziej:2003gp}. A preliminary study of the sensitivity of $Wtb$ anomalous couplings on the single top quark production cross-section in $e^-p$ collision for TESLA + HERA and LHC+CLIC energies has been performed in  \cite{Atag:2001qf}. 
\par Recently a deep inelastic electron-nucleon scattering facility is proposed at the LHC, known as LHeC. It is proposed that an electron beam of 60 GeV will collide with 7 TeV proton beam simultaneous   to the existing proton-proton collision  experiments at the LHC \cite{AbelleiraFernandez:2012cc,AbelleiraFernandez:2012ni,Bruening:2013bga}.
The LHeC is expected to test the rich electroweak physics with precision. There has been some work on the physics goals of the collider \cite{AbelleiraFernandez:2012cc,AbelleiraFernandez:2012ni,Bruening:2013bga, Senol:2012fc,Biswal:2012mp,Han:2009pe}.
The working group  involved in  the synergy between the LHC and the LHeC brought out an excellent report showing the inter-dependencies of the physics reach and goals of both these colliders \cite{AbelleiraFernandez:2012ty} . The LHeC is going to provide an unprecedented platform for studying the single top quark production as this has an advantage over the LHC and the Tevatron in terms of providing  (a) a clean environment with suppressed background from strong interaction initiated processes, and (b) a kinematic reach for lepton-nucleon scattering at c.m. energy around 1.3 TeV \cite{Kumar:2015jna,Andreazza:2015bja,Sarmiento-Alvarado:2014eha,Agashe:2013hma,Bouzas:2013jha}.

Thus it is worthwhile to study the single top quark production and probe the $Wtb$ anomalous couplings  at the LHeC. 
\par In Sec. \ref{stoplhec} we analyse and study the single anti-top quark production and potential backgrounds, their yield, choice of selection  cuts and kinematic distributions at the LHeC. We introduce kinematic asymmetries as estimators in Sec. \ref{chisqanalysis}, provide the exclusion contours based on bin analysis of  distributions involving kinematic observables and finally using the method of optimal variables we give error correlation matrices and exclusion contours with 1\% luminosity uncertainty. We discuss the impact of the luminosity uncertainty on the measurement of the couplings and their correlations. The summary and analysis of our observations are given in Sec. \ref{summary}.
\section{Single anti-top quark production}
\label{stoplhec}
In hadron colliders, the SM single top quark production at leading order is studied through three disparate non-interfering  modes via $s$-, $t$- and $Wt$- channels, respectively and details can be found in \cite{Dutta:2012ai}. The $t$- channel through charge current (CC) interactions dominates over all the other production mechanism. In the LHeC we can study the single top quark production only through $t$ channel process $e^{-} \bar b \to \nu_e \bar t + X$ as  shown in Figure \ref{fig:stop}. 
In sharp contrast to the LHC the absence of pile-up and underlying event effects  at the LHeC, high rates of single anti-top production  is expected to provide a better insight on $Wtb$ anomalous couplings. The sensitivity of the $Wtb$ couplings are also investigated through the sub-dominant  associated $tW$ production  in references \cite{Cakir:2013qw,Cakir:2009rq}. 
\par We have implemented $Wtb$ effective couplings corresponding to both chiral vector and tensor structures given by the Lagrangian  \eqref{lagwtb} in MadGraph/MadEvent \cite{Alwall:2011uj} using  FeynRules \cite{Alloul:2013fw}.  The partonic cross sections are convoluted with CTEQ6L1 parton distribution functions (PDF) keeping factorization and renormalization scale $\mu_F = \mu_R = m_t = 172.5$ GeV. The mass of $b$-quark $m_b=4.7$ GeV and $W^\pm$ boson $m_W$= 80.399 GeV, assuming the SM value for $\left|V_{tb}\right|\, f_1^L$ = 1.
\par  The total top decay width which is  is one of the  fundamental property of top physics is measured with precision from the partial decay width $\Gamma (t\to W\,b)$ in the $t$ channel of the single top quark production.  The effect of anomalous $Wtb$ couplings in evaluating the decay width of the anti-top quark  is consistently taken into account throughout our analysis for the  signal cross-section.

\par Considering the five flavor constituents of proton we study  the $2\to 2$ process $e^{-}  p \to \nu_e \bar t + X$ and probe the accuracy with which the anomalous couplings can be measured. The variation of the cross-section of the single top production in SM is studied with respect to the center of mass energy and electron energy in Figure \ref{xsecvar} and we are in agreement  with the earlier results given in \cite{Atag:2001qf}. We also show the effect of taking 80\% beam polarization for electron, which results in the enhancement of the SM single top production cross section as the cross-section scales as $(1+P_{e^-})$, $P_{e^-}$ being the degree of polarization of the electron.
\par We also depict the varying contribution of $2\to 3$ process $e^-p\to \bar t \,\nu_e\, b$ from the four flavor proton where the gluon splits into $b,\,\bar b$ and $\bar b$ participates in the interaction while $b$ quark is produced in final state as a spectator quark. This process is however suppressed in comparison to the $2\to 2$ process $e^-p\to \bar t \,\nu_e$. This signal can be vetoed out by demanding the exclusion of two $b$ jets. We do not consider this process for our analysis.
\par For the rest of the analysis we compute all cross-sections for the proposed LHeC with $E_{e^-}=60$ GeV and $E_p=7$ TeV as per recommendations given in the LHeC conceptual design report  \cite{AbelleiraFernandez:2012cc}. The total events are estimated with an integrated luminosity $L$ = 100 fb$^{-1}$. 
\par  The new physics effect can arise either at  the production vertex of the anti-top in the process $e^- p\to \bar t \nu_e \to \bar b W^- \nu_e $ or at the decay vertex. Figure \ref{npxsecvar}  depicts the interplay of the interference terms  for the left handed current and shows the  variation of the cross section with respect to the variation in the anomalous couplings.  
\par  The stronger dependence of the cross section on the anomalous coupling $\Delta f_1^L$ is because of the identical Lorentz structure associated with the SM and $\Delta f_1^L$ and accordingly the constructive (destructive) interference becomes pronounced for positive (negative) $\Delta \left\vert f_1^L\right\vert$.  Therefore the cross-section of left handed vector current mediated process  varies as  $\left[(1+\Delta f_1^L) \left\vert V_{tb}\right\vert\right]^2$.  On the other hand,   the right handed current mediated processes  vary as $\left\vert f_i^R\right\vert^2$ for $i=1,2$  and are therefore  sub-dominant even in the presence of  large $\left\vert f_i^R\right\vert$ because of the non-SM structure of the current.
\par We estimate and study  the $W^-$ helicity distributions arising from  NP effects. The $W$ polarization distribution distinguishes the contribution of anomalous couplings. We study the behaviour of the helicity fractions of the $W^-$ in terms of ratios of the number of events ${\cal F}_-=N_-/N$, ${\cal F}_+=N_+/N$ and ${\cal F}_0=N_0/N$ where $N_-$, $N+$ and $N_0$ are the left, right and longitudinally polarized $W^-$ events and $N=N_++N_-+N_0$. 

We vary the coupling and study its effect through the variation on these ratios in Figure \ref{hel_ratios}. We observe that
\begin{enumerate}
\item[(a)] The ${\cal F}_-$ and ${\cal F}_+$ corresponding to the positive and negative polarized $W$'s show opposite trend with the variation of all effective couplings except $\left\vert v_{tb}\right\vert\Delta f^L_1$.

 \item[(b)] The helicity fractions ${\cal F}_i$ associated with the left handed  tensor current is most sensitive as it interferes with the SM and has a larger momentum  dependence.  Right handed vector chiral current shows an appreciable sensitivity {\it w.r.t. } ${\cal F}_i$ helicity distribution. 
\par The  helicity fractions ${\cal F}_-$ and ${\cal F}_0$ are also sensitive to the change in the coefficient of the right handed tensor current.
\end{enumerate}  
\begin{onecolumn}
\begin{table}[htb]\footnotesize
\centering
\begin{tabular*}{\textwidth}{c|@{\extracolsep{\fill}} cccccc}\hline\hline
%\centering
%\begin{tabular}{|c|c|c|c|c|c|c|c|} \hline\hline
No. & Background   & ${p_T}_{j,b} \ge 20$ GeV    &$\Delta \Phi_{\not\!E,j} \ge 0.4$&$\left\vert m_{j_1j_2}-m_W\right\vert \le 22\, {\rm GeV}$&$\sigma_{\rm eff.}$ \\
    & Process      &$\left\vert\eta_j\right\vert \le 5$,$\left\vert\eta_b\right\vert \le 2.5$  & $\Delta \Phi_{\not\! E,b} \ge 0.4$ & &\\
    &              &$\Delta R_{j,b/j}\ge 0.4$   &  &  & \\
    &              &$\not \!\! E_T\ge 25$       &  &  & &\\ \hline \hline
1   & $e^-p\to \nu_eW^-\bar b$        & $7.5 \times 10^{-3}$      & $6.8 \times 10^{-3}$   & $4.5 \times 10^{-3}$                 & $2.7 \times 10^{-3}$  \\
    & without anti-top line           &                          &                       &                      &\\\hline
2   & $ e^- p \to \nu_e  j j j$       & $4.2 \times 10^0$          & $3.6 \times 10^0$       & $2.4 \times 10^{0}$&  $7.2 \times 10^{-2}$  \\\hline
3   & $  e^- p \to \nu_e c j j$       & $1.5 \times 10^0$          & $1.2 \times 10^{0}$     & $8.6 \times 10^{-1}$& $8.6 \times 10^{-2}$   \\
    & \& $e^- p \to \nu_e \bar c j j$& &&&\\\hline
4   & $e^- p \to \nu_e c \bar c j$    & $5.8 \times 10^{-2}$       & $5.0 \times 10^{-2}$    & $3.2 \times 10^{-2}$ & $6.7 \times 10^{-3}$  \\ \hline
5   & $e^- p \to \nu_e   b \bar b j$  & $2.5 \times 10^{-2}$      & $2.2 \times 10^{-2}$   & $5.6 \times 10^{-3}$& $1.3 \times 10^{-3}$ \\ \hline
6   & $ e^- p \to \bar c\nu_e $       & $2.5 \times 10^{-2}$      & $2.2 \times 10^{-2}$   &$1.5 \times 10^{-2}$    & $1.5 \times 10^{-4}$    \\ 
    & $(\bar c\to W^- \bar s)$        &&&&\\
\hline\hline
\end{tabular*}
\caption{\small \em{ Cross-section of all background processes in $pb$ for the hadronic channel  with selection cuts. The effective background cross-section $\sigma_{\rm eff.}$ is computed in the fifth column by multiplying  $b/\bar b$ tagging efficiency and/ or faking probability  1/10 and 1/100  corresponding to final state charm /anti-charm and  light jets $j\equiv u,\bar u, d,\bar d,s,\bar s,g$, respectively.    }}
\label{Bkgnd_had}
%\end{table}
% \begin{table}[htb]\footnotesize
%\centering
%\begin{tabular}{|c||c|c|c|c|c|c|c|c|}\hline \hline 
%\centering
\begin{tabular*}{\textwidth}{c|@{\extracolsep{\fill}} ccccc}\hline\hline
%\centering
%\begin{tabular}{|c|c|c|c|c|c|c|c|} \hline\hline
Event Selection   & ${p_T}_{j,b} \ge 20$ GeV    &$\Delta \Phi_{\not\!E,j} \ge 0.4$&$\left\vert m_{j_1j_2}-m_W\right\vert \le 22\, {\rm GeV}$ & Fiducial& $S/\sqrt{S+B}$\\
                  & $\left\vert\eta_j\right\vert \le 5$,$\left\vert\eta_b\right\vert \le 2.5$  & $\Delta \Phi_{\not\! E,b} \ge 0.4$ &  & Efficiency&\\
                  & $\Delta R_{j,b/j}\ge 0.4$   &  &  & &\\
                  & $\not \!\! E_T\ge 25$       &  &  & &\\ \hline \hline
$SM$                                          & $3.2 \times 10^4$    & $2.3 \times 10^4$   & $2.2 \times 10^4$   & 66.7 $\%$ & -- \\
SM+$\sum_i {\rm Bkg}_i$                       & $6.5 \times 10^4$    & $5.0 \times 10^4$   & $4.0 \times 10^4$   & 61.5 \% & \\
$\left|V_{tb}\right|\Delta f_{1}^{L}=.5$      & $7.3 \times 10^4$    & $5.0 \times 10^4$   & $5.0 \times 10^4$   & 68.0 $\%$ & 1.92 \\
$f_{1}^{R}=.5$                                & $4.6 \times 10^4$    & $3.2 \times 10^4$   & $3.2 \times 10^4$   & 69.7 $\%$ & 1.43 \\
$f_{2}^{L}=.5$                                & $4.9 \times 10^4$    & $3.6 \times 10^4$   & $3.6 \times 10^4$   & 73.2 $\%$ & 1.55 \\ 
$f_{2}^{L}=-.5$                               & $3.4 \times 10^4$    & $2.3 \times 10^4$   & $2.3 \times 10^4$   & 69.6 $\%$ & 1.40 \\ 
$f_{2}^{R}=.5$                                & $5.7 \times 10^4$    & $4.1 \times 10^4$   & $4.1 \times 10^4$   & 72.3 $\%$ & 1.69 \\ 
\hline\hline
\end{tabular*}
\caption{\small \em{ Yield   with selection cuts  in the hadronic channel corresponding to the chosen anomalous coupling value of 0.5  at integrated  luminosity $L$ = 100 fb$^{-1}$. The yield corresponding to  SM+$ \sum_i {\rm Bkg}_i$ signify the total cumulative events of SM and all backgrounds  after taking into account the  $b,\bar b$ faking/tagging efficiency. Yields corresponding to all anomalous couplings include the SM top background. }}
\label{yield_had}
%\centering
\begin{tabular*}{\textwidth}{c|@{\extracolsep{\fill}} lccc}\hline\hline
%\begin{tabular}{|c|c|c|c|c|c|c|c|} \hline\hline
No. &  Background     &${p_T}_{j,b,l} \ge 20$ GeV, $\Delta R_{j,b/j}\ge 0.4 $, $\not \!\! E_T\ge 25$ &$\Delta \Phi_{\not\!E,j} \ge 0.4$& $\sigma_{\rm eff.}$\\
    &  Process        &$ \left\vert\eta_j\right\vert \le 5,\, \left\vert\eta_{b,l}\right\vert \le 2.5$&$\Delta \Phi_{\not\! E,b} \ge 0.4$&\\ 
    &                 &                                                                               &$\Delta \Phi_{\not\! E,l} \ge 0.4$&\\ \hline \hline
1   &  $e^-p\to l^-\bar\nu_l\nu_e j$      & $1.5 \times 10^{-1}$    & $1.4 \times 10^{-1}$ &$1.4 \times 10^{-3}$\\\hline
2   &  $e^-p\to l^-\bar\nu_l\nu_e c$      & $6.6 \times 10^{-3}$    & $6.1 \times 10^{-3}$ &$6.1 \times 10^{-4}$\\
    &  \& $ e^-p\to l^-\bar\nu_l\nu_e \bar c$& &&\\ \hline
3   &  $e^-p\to l^-\bar\nu_l\nu_e b $     & $3.6 \times 10^{-3}$    & $3.2 \times 10^{-3}$  &$1.9 \times 10^{-3}$\\
    &  \& $e^-p\to l^-\bar\nu_l\nu_e \bar b $ &&&\\ 
    &  Without top line &&\\ \hline
4   &  $e^-p\to e^- l^-\bar\nu_l  c $     & $1.5 \times 10^{-2}$    & $6.9 \times 10^{-3}$&$6.9 \times 10^{-4}$ \\ \hline
5   &  $e^-p\to e^- l^-\bar\nu_l  j $     & $1.2 \times 10^{-1}$    & $5.5 \times 10^{-2}$ &$5.5 \times 10^{-4}$\\ \hline\hline
\end{tabular*}
\caption{\small \em{ 
Cross-section of all background processes in $pb$ for the leptonic channel  with selection cuts.
The effective background cross-section $\sigma_{\rm eff.}$ is computed in the fourth column by multiplying  $b/\bar b$ tagging efficiency and/ or faking probability  1/10 and 1/100  corresponding to final state charm /anti-charm and  light jets $j\equiv u,\bar u, d,\bar d,s,\bar s,g$, respectively.
  The background processes with two charged leptons are taken into consideration where one get lost in the beam pipe. }}
\label{Bkgnd_lep}
\begin{tabular*}{\textwidth}{c|@{\extracolsep{\fill}} cccccc}\hline\hline
%\centering
%\begin{tabular}{|c|c|c|c|c|c|c|c|} \hline\hline
Event Selection   & ${p_T}_{j,b} \ge 20$ GeV    &$\Delta \Phi_{\not\!E,j} \ge 0.4$  & Fiducial&$S/\sqrt{S+B}$ \\
                  & $\left\vert\eta_j\right\vert \le 5$,$\left\vert\eta_b\right\vert \le 2.5$  & $\Delta \Phi_{\not\! E,b} \ge 0.4$  & Efficiency&\\
                  & $\Delta R_{j,b/j}\ge 0.4$   & $\Delta \Phi_{\not\! E,l} \ge 0.4$ &   &\\
                  & $\not \!\! E_T\ge 25$       &  &   &&\\ \hline \hline
SM                                            & $1.2 \times 10^4$    & $1.1 \times 10^4$   & 92.0 $\%$ & --\\
SM+$\sum_i {\rm Bkg}_i$                       & $1.3 \times 10^4$    & $1.2 \times 10^4$   & 92.0 $\%$ & --\\
$\left|V_{tb}\right|\Delta f_{1}^{L}=.5$      & $2.7 \times 10^4$    & $2.5 \times 10^4$   & 92.6 $\%$ & 1.55\\
$f_{1}^{R}=.5$                                & $1.7 \times 10^4$    & $1.6 \times 10^4$   & 94.1 $\%$ & 1.23\\
$f_{2}^{L}=.5$                                & $1.9 \times 10^4$    & $1.7 \times 10^4$   & 89.5 $\%$ & 1.27 \\ 
$f_{2}^{L}=-.5$                               & $1.1 \times 10^4$    & $1.0 \times 10^4$   & 90.9 $\%$ & 0.95\\ 
$f_{2}^{R}=.5$                                & $2.2 \times 10^4$    & $2.0 \times 10^4$   & 90.9 $\%$ & 1.38\\ 
\hline\hline
\end{tabular*}
\caption{\small \em{Yield with selection cuts  in the leptonic channel corresponding to the chosen anomalous coupling value of 0.5  at integrated  luminosity $L$ = 100 fb$^{-1}$. The yield corresponding to  SM+$ \sum_i {\rm Bkg}_i$ signify the total cumulative events of SM and all backgrounds  after taking into account the appropriate  $b,\bar b$ faking/tagging efficiency.}}
\label{yield_lep}
\end{table}	
\end{onecolumn}
\begin{twocolumn}

\par The  helicity fractions are recently measured in the top quark pair events decaying leptonically and semi-leptonically with $\sqrt{s}=8$ TeV at CMS detector in LHC with an integrated luminosity of 5 $fb^{-1}$ \cite{Chatrchyan:2013jna}.  Constraints  obtained on $F_-$ and $F_0$ are found to be consistent with SM but observations has left $F_+$ unconstrained.

\par Helicity fractions are studied through the recons- tructed tops/ anti-tops in the experiment. Therefore sensitivity of these helicity fractions are subjected to systematic uncertainties arising from the re-construction algorithm efficiency and the determination of the angular distribution of all the decay products of the top/ anti-top. However one can overcome the above shortcomings with large statistics {\it e.g} in LHC and improving the reconstruction of the most extreme bins in the angular distribution \cite{AguilarSaavedra:2007rs}. Moreover, it is better studied in  hadron colliders where tops/ anti-tops are dominantly produced through strong interaction vertices for which  the $Wtb$ anomalous coupling would  only depend on the decay vertices of  tops/ anti-tops.
\par In this article, we proceed to extract more information on the sensitivity of anomalous couplings through one dimensional distribution of kinematic variables in the following section.

\par  Finally we  analyse the anti-top through  the hadronic and  leptonic decay modes of $W$'s. Henceforth, we have  multiplied the cross-section (for processes having  $b$ or $\bar b$ as its final state)   with $b,\,\bar b$ tagging efficiency $\epsilon_b\,=$  0.6.

\subsection{Sensitivity in the Hadronic Mode}
In order to study the sensitivity of the anomalous couplings introduced in equation \eqref{lagwtb}, we  examine the process $e^- p\to \bar t \nu_e, (\bar t \to W^{-} \bar b, W^{-} \to j j)$, $j\equiv \bar u,d,\bar c,s$ at the LHeC and its potential backgrounds. We impose standard selection cuts as follows
\begin{enumerate}[(i)]\label{cuts}
\item Minimum  transverse momentum for jets, $\bar b$-antiquark  $p_{T_{b,j}} \ge 20$ GeV, $p_{T_{j,\bar l}} \ge 25$ GeV and minimum missing transverse energy $\slashed E_T \ge 25$ GeV.  
\item The pseudo-rapidity region for leptons and $\bar b$-antiquark $ \left\vert\eta_{\bar b,l} \right\vert $ is taken to be $\le 2.5$, however for jets $\left\vert \eta_j\right\vert \le 5$.   
 
\item Isolation cuts for  lighter, heavy quarks and lepton require $\Delta R_{ij} \ge 0.4 $ where $i, \,j\equiv$ leptons, jets and $\bar b$ anti-quark.      
\end{enumerate}

In addition, we impose the following cuts to reduce the background
\begin{enumerate}\label{addcuts}
\item[(iv)] The difference of azimuthal angle between missing energy $\slashed E_T$ and jets, leptons, $\bar b$-antiquark should be $\Delta \phi \ge 0.4$.  
 \item[(v)] To further reduce the background in the hadronic channel we reconstruct $W^-$  from di-jets  assuming the jet energy resolution $\approx\frac{\sigma}{E}=\frac{0.6}{\sqrt {E}}$. In this setup the di-jet invariant mass resolution around the $W^-$ mass is approximately  7$\%$. Thus a mass window around 28\% (4 times of this resolution at $2\sigma$ level) of the $W$ mass $\approx 22$ GeV is taken into consideration and hence di-jet invariant mass is allowed to satisfy $\left\vert m_{j_1\,j_2}-m_W\right\vert \le 22$ GeV.  
\end{enumerate}
The cross-section of the background processes and the effect of these  selection cuts are given in the Table \ref{Bkgnd_had}. The effective cross-section given in the fifth column is calculated after multiplying the $b\bar b$ tagging efficiency of 0.6.
The $b$ or $\bar b$ faking  probability is taken to be  $1/100$ for $u,d,s$ quarks, antiquarks and $1/10$ for $c,\bar c$ quarks.

We observe that  
\begin{enumerate}[(a)]
\item The dominant background process is  $e^- p \to \nu_e c/\bar c (j j)$ where $j\equiv u,\bar u, d,\bar d, s,\bar s,  g$. The effective irreducible cross-section of the this background after imposing all cuts is $\approx 0.1$ pb. The other dominant background is 
$e^- p \to \nu_e j\,j\, j$ which along with the first one constitute almost 94\% of the total irreducible background 169 fb.

\item the cross-section of $e^{-} p \to \nu_e W^- \bar b$ is   dominated by diagrams wherein the $W^- \bar b$ is generated  from anti-top quarks.  However, after multiplying with the appropriate branching ratio for the hadronic mode of $W^-$ the cross-section is reduced to the order of $10^{-3}$ pb. We have also found that the  potential background due to mis-tagging of one of the double $b,\bar b$ events arising from the process to $e^- p \to \nu_e j b \bar b$ is negligibly small. 
\end{enumerate}
\end{twocolumn}

To probe the effect of these cuts on the yield, we  study all kinematic distributions   in SM, other non-top backgrounds and compare them with contributions from new physics cases with the representative value of the effective  coupling at 0.5.
The analysis is summarized in Table \ref{yield_had} and the overall fiducial efficiencies of additional cuts are presented.   The significance $S/\sqrt{S+B}$ give the sensitivity of the cross-section   corresponding to  these representative values. The yield of background processes
mentioned in Table \ref{Bkgnd_had} is computed by taking the appropriate weight factor due to mis-tagging or tagging of light quarks and $\bar b$ quark respectively. 
\par
\begin{figure*}[htb]
  \centering
  \begin{tabular}{cc}
  \includegraphics[width=0.5\textwidth,clip]{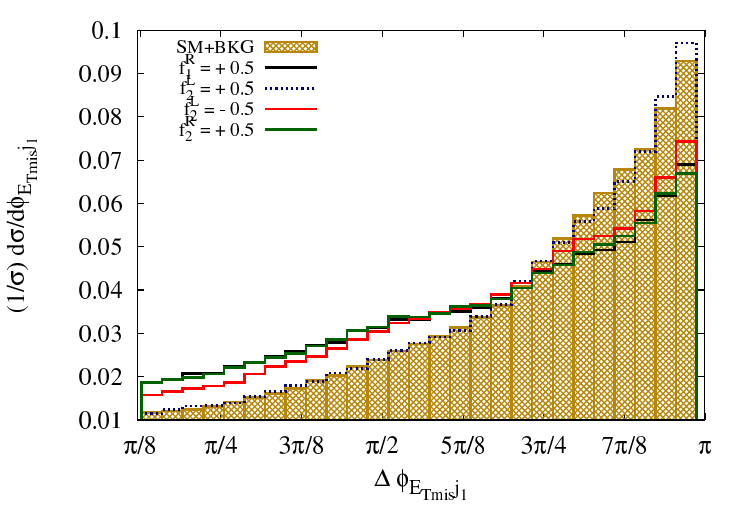} &
  \includegraphics[width=0.5\textwidth,clip]{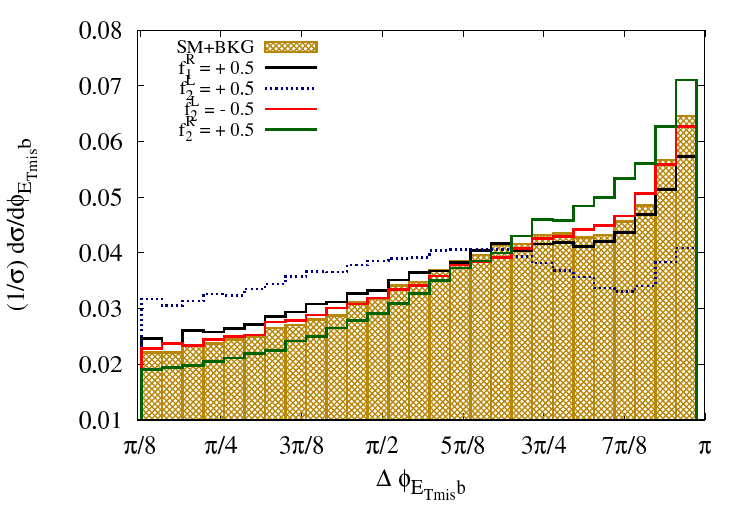} \\
  \includegraphics[width=0.5\textwidth,clip]{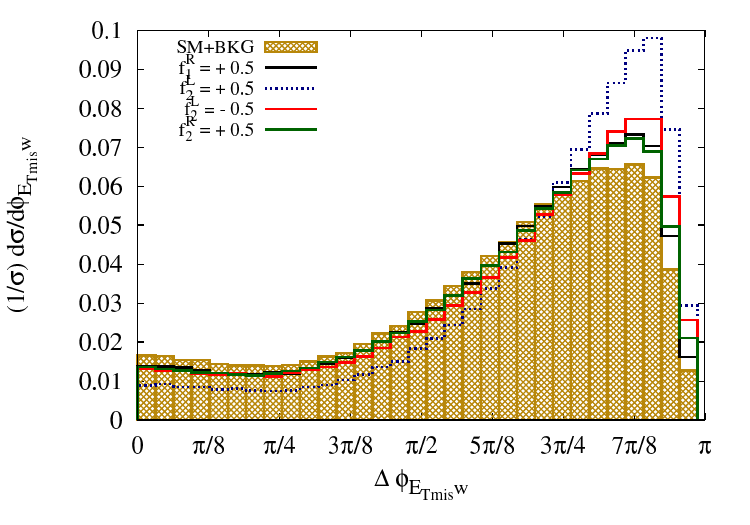} &
  \includegraphics[width=0.5\textwidth,clip]{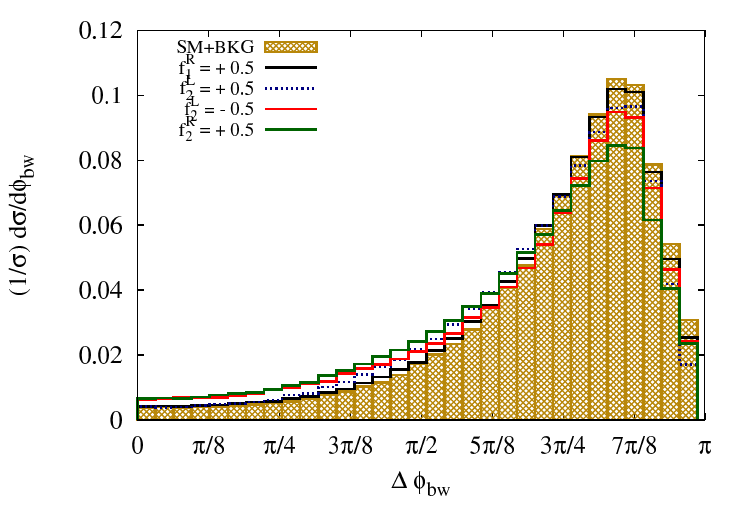} \\
 \includegraphics[width=0.5\textwidth,clip]{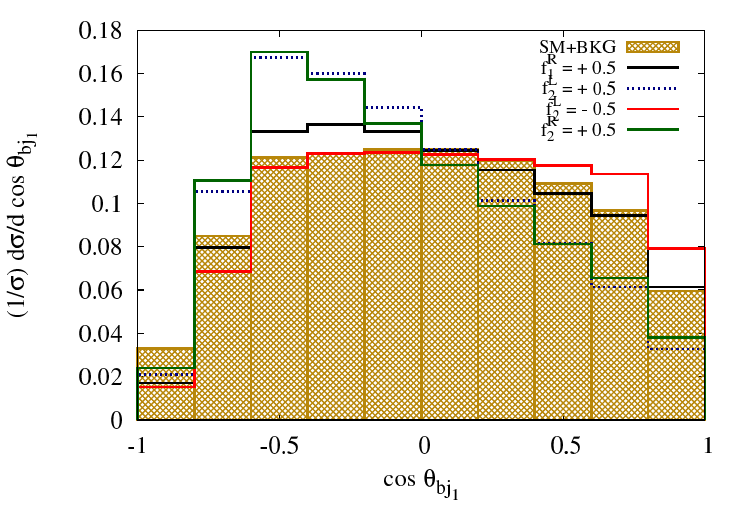} &
 \includegraphics[width=0.5\textwidth,clip]{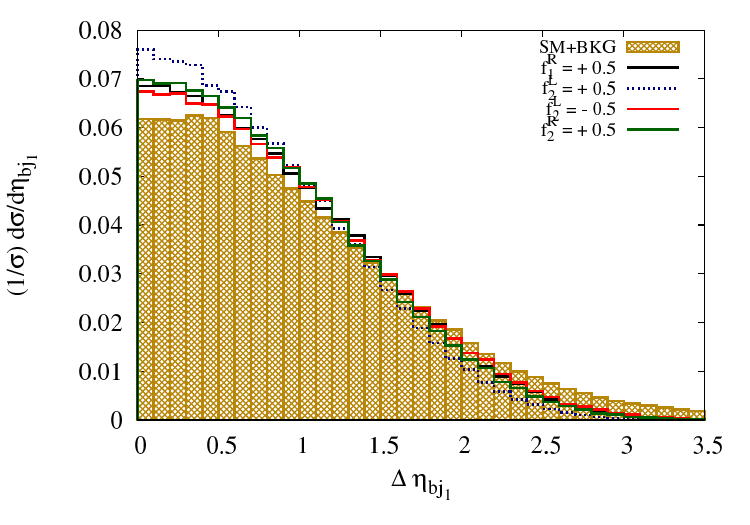}  
  \end{tabular}
  \caption{\small \em{ Normalized distributions of $\Delta\phi_{\slashed E_T\, j_1}$,  $\Delta\phi_{\slashed E_T\, \bar b}$, $\Delta\phi_{\slashed E_T\, W}$, $\Delta\phi_{b \, W}$, $\cos\theta_{\bar b\, j_1}$ and  $\Delta\eta_{\bar b\, j_1}$  for hadronic decay mode of $W^-$, corresponding to SM and an anomalous coupling of 0.5. Here $j_1$ is the highest $p_T$ jet. The normalized distributions corresponding to $\left\vert V_{tb}\right\vert \Delta f_1^L=\pm 0.5$ is identical to that of SM. All kinematic observables are measured in lab frame.}}
\label{metXhad}
\end{figure*}

 The characteristics of the highest $p_T$ jet $j_1$, the final state $\bar b$ and the missing transverse energy $\slashed E_T$ are likely to bear the signature of the $Wtb$ couplings at the production/ decay vertex. We reconstruct the $W^-$ from jets at the final states to study the  azimuthal angle separation between   $W-$ and $\bar b$ and missing energy $\slashed E_T$. We study   one dimensional distributions of azimuthal angle (angle between the planes) $\Delta\phi_{\slashed E_T,\,j_1}$, \,\, $\Delta\phi_{\slashed E_T,\,\bar b}$, \,\, $\Delta\phi_{\slashed E_T,\,W} $ and $\Delta\phi_{\bar b,\,W}$ along with the $\cos\theta_{\bar bj_1}$ and $\Delta\eta_{\bar bj_1}$, where all angles are defined in the lab frame.  Figure \ref{metXhad} exhibit these distributions. To study the distribution profile and shape variation, all histograms are normalized to unity to and are drawn for a anomalous coupling representative value 0.5. The normalized distributions corresponding to $\left\vert V_{tb}\right\vert \Delta f_1^L=\pm 0.5$ is identical to that of SM. However, on consideration of backgrounds the  distribution profile of kinematical variable generated from $\left\vert V_{tb}\right\vert\Delta f_1^L=\pm 0.5$ shows  distortion when compared to that of pure SM. The SM+$\sum {\rm Bkg}_i$ distributions are drawn after summing  the bin-wise contribution from each background process with appropriate factor as mentioned earlier.

In most of the distributions the new physics couplings play a significant role and clear distinction has been seen in profile with respect to combined effect SM and backgrounds. 
We observe from Figure \ref{metXhad} that the contribution of left and right handed tensorial Lorentz structures are distinguishable in most distributions. The distributions corresponding to (a) azimuthal angle  between missing energy and highest $p_T$ jet $j_1$ and (b)  cosine of the angle between massive $b$ quark and $j_1$ show a noticeable difference in the profile with respect to the right handed tensor chiral current.
 
\subsection{Sensitivity in the Leptonic Mode}
Similarly we  study the yield of the leptonic decay mode of $W^-$ through the process $e^- p \to \bar t \nu_e, \,\, \left(\bar t \to W^{-} \bar b,\,\, W^{-} \to l^- \bar \nu_l\right)$, $l^-\equiv e^-,\mu^-$ at the LHeC. We impose  the standard selection cuts are same as those given in \ref{cuts}. The effects of these selection cuts are given in Table \ref{Bkgnd_lep}. The effective cross-section is given in the fourth column of this table. In general all backgrounds processes are sub-dominant. Reading this Table \ref{Bkgnd_lep}, we observe that
\begin{enumerate}
\item[(a)] processes with a charged lepton, $\slash \!\!\! \! E_T$ and light jets, where the light jets can fake a $b$ jet of the signal becomes negligibly small once they are screened through the selection cuts and  multiplied by the appropriate faking probability factor.
\item[(b)] background processes with two charged leptons where one of them vanishes in the beam pipe  is negligible   after the imposition of the selection cuts. 
\end{enumerate}
\par The fiducial efficiencies due to the additional cuts are computed for the representative value of couplings at $\pm 0.5$   corresponding to the coefficient of the different chiral and Lorentz structures as given in \eqref{lagwtb}. They are shown along with the significance in Table \ref{yield_lep}. 
%It is interesting to note that the significance $S/\sqrt{S+B}$  is reduced to an large extent 
\par In the leptonic mode the final state charged lepton  along with $\bar b$ shows the characteristic features of the anomalous couplings. Further we study  the sensitivity of the couplings through  one dimensional distributions corresponding to azimuthal angle $\Delta\phi_{\slashed E_T,\,l_1}$, \,\, $\Delta\phi_{\slashed E_T,\,\bar b}$, along with the polar angle $ \cos\theta_{\bar bl_1}$ and difference of pseudo-rapidity $\Delta\eta_{\bar bl_1}$ between $\bar b$ and   the charged lepton with highest $p_T$ designated as $l_1$. Figure \ref{metXlep} depict these distributions. As mentioned before all normalized distributions corresponding to $\left\vert V_{tb}\right\vert \Delta f_1^L=\pm 0.5$ are identical to that of  SM single top production. We observe that $f_2^L$ shows a distinguishable profile over others. However, the distribution $\Delta \phi_{\slashed E_T\,l}$ is sensitive to all anomalous couplings. 
\section{Estimators and $\chi^2$ analysis}
\label{chisqanalysis}
\subsection{Angular Asymmetries from Histograms}
\label{estimators}
We construct the asymmetry from the  distribution of  kinematic observables  in both the hadronic and leptonic modes. These asymmetries can  be  sensitive discriminators  to distinguish  the contribution from the different Lorentz structure due to their characteristic momentum dependence. We study the angular asymmetries with respect to the polar angle $\cos\theta_{ij}$, rapidity difference $\Delta \eta_{ij}$ and azimuthal angle difference $\Delta\phi_{ij}$, where $i,j$ may be any partons (including $\bar b$-antiquark), charged lepton or missing energy. The associated asymmetries $A_{\theta_{ij}}$,  $A_{\Delta \eta_{ij}}$ and $A_{\Delta\Phi_{ij}}$ are defined as
\begin{eqnarray}
&A_{\theta_{ij}}&=\,\frac{ N_+^A(\cos\theta_{ij} > 0) - N_-^A(\cos\theta_{ij} < 0) }{ N_+^A(\cos\theta_{ij} > 0) + N_-^A(\cos\theta_{ij} < 0) } \label{atheta} \\
&A_{\Delta \eta_{ij}}&=\,\frac{ N_+^A(\Delta \eta_{ij} > 0) - N_-^A(\Delta \eta_{ij} < 0) }{ N_+^A(\Delta \eta_{ij} > 0) + N_-^A(\Delta \eta_{ij} < 0) } \label{ady} \\
&A_{\Delta\Phi_{ij}} &=\, \frac{N_+^A\left( \Delta\phi_{ij}>\tfrac{\pi}{2}\right)-N_-^A\left( \Delta\phi_{ij}<\tfrac{\pi}{2}\right)}{N_+^A\left( \Delta\phi_{ij}>\tfrac{\pi}{2}\right)+N_-^A\left( \Delta\phi_{ij}<\tfrac{\pi}{2}\right)} \label{aphi}
\label{adphi}
\end{eqnarray}
with $0\leq\Delta\phi_{ij}\leq \pi$. The asymmetry $A_{\alpha}$ and its statistical error for $N_{+}^A$ and $N_{-}^A$ events where $N=\left(N_{+}^A+N_{-}^A\right)= L \cdot\sigma$ is calculated by using the following definition based on binomial distribution :
\begin{eqnarray}
A_{\alpha} &=& a \pm \sigma_a, \qquad \text{where}\,\,a = \frac{N_{+}^A - N_{-}^A}{N_{+}^A+ N_{-}^A}\,\, \text{and}\nonumber \\
 &&  \sigma_a = \sqrt{\frac{1-a^2}{L\cdot\sigma}};\quad \left(\alpha=\cos\theta_{ij},\Delta \eta_{ij},\Delta \Phi_{ij} \right) \label{aerr}
\end{eqnarray}
Here $\sigma \equiv\sigma (e^-p\to \bar t \nu, \, \bar t \to W^- \bar b) \times BR( W^- \to jj/\, l^-\bar\nu)\times \epsilon_b $  is the total  cross-section in the respective channel after imposing selection cuts and $\epsilon_b=0.6$ is the $b/\bar b$ tagging efficiency.

Based on the one dimensional histograms  given in Figures \ref{metXhad} and \ref{metXlep}, we look for the asymmetry within a distribution generated due to the interplay of the SM, Background channels  and a given anomalous coupling for two distinct hadronic and leptonic modes of $W^-$ decay.  Any large  deviation from the combined asymmetry due to SM and background processes    would then imply that the associated kinematic observable is an optimal variable in determining the sensitivity of the given anomalous coupling. We provide these asymmetries constructed from the distributions in Table \ref{asym_had}  and \ref{asym_lep} for the hadronic and leptonic channels, respectively a representative value of the anomalous coupling 0.5. Any asymmetry with respect to distributions corresponding to $\left\vert V_{tb}\right\vert\Delta f_1^L$  is identical to the one in  SM.
\par Asymmetries shown in Tables  \ref{asym_had} and \ref{asym_lep} are good estimators for preliminary studies. They give a handle for judging the ability of the measured observable to distinguish the contribution from an anomalous term in the Lagrangian. We observe  in Table \ref{asym_had} that the couplings are sensitive in magnitude as well as sign of  the asymmetry generated by $\cos\theta_{b\,j_1}$ distribution. But they may not be sensitive enough for the couplings which are  one order of magnitude smaller than the representative value. In fact  the whole distribution is essentially divided into two halves which correspond to only two bins with large bin-width. 
\subsection{Exclusion contours from bin analysis}
\label{bin analysis}
In this subsection the sensitivity of couplings are obtained through $\chi^2$ analysis, where we compute the sum of the variance of events over all bins. 
Thus  more bin information is likely to yield better sensitivity than the asymmetries which are generated essentially by dividing the whole distribution  into two equal bins.
\par To make the analysis more effective we switch on two effective anomalous couplings at a time with SM and all possible background processes with same final states.  The $\chi^2$ becomes a function of two effective anomalous couplings $f_i$, $f_j$ and defined as
\begin{eqnarray}
 \chi^2\left(f_i,f_j\right) 
 &=&\sum_{k=1}^N \,
  \left(\, \frac{{\cal N}^{\rm exp}_k - {\cal N}^{\rm
   th}_k\left(f_i,f_j\right) }{\delta{\cal N}^{\rm exp}_k} \,
  \right)^2 \label{chidef} 
\end{eqnarray}
where ${\cal N}_k^{\rm th}\left(f_i,f_j\right)$ and  ${\cal N}_k^{\rm exp}$  are the total number of events predicted by the theory involving $f_i,\, f_j$ and   measured in the experiment for the  $k^{\rm th}$ bin. $\delta {\cal N}_k^{\rm exp}$ is the combined statistical and systematic error  $\delta_{\rm  sys}$
 in measuring the events for the  $k^{\rm th}$ bin.
 If all the coefficients $f_i$'s are small, then the  experimental
result in the $k^{\rm th}$ bin should be approximated by the SM and background prediction as 
%\vskip -1.cm
\begin{eqnarray}
 {\cal N}^{\rm exp}_k \approx 
  {\cal N}^{\rm SM}_k+\sum_i {\cal N}^{\rm Bkg_i}_k= {\cal N}^{\rm SM+\sum_i Bkg_i}_k.
  \label{expsm}
\end{eqnarray}
The error  $\delta {\cal N}_k^{\rm SM}$ can be defined as
\begin{eqnarray}
\delta {\cal N}_k^{\rm SM+\sum Bkg_i} = \sqrt{{\cal N}_k^{\rm SM+\sum_i Bkg_i} \left( 1 + \delta_{\rm  sys}^2 \,\,{\cal N}_k^{\rm SM+\sum_i Bkg_i} \right)}.
\end{eqnarray}
\noindent The $\chi^2$ analysis due to un-correlated   systematic uncertainties  is studied for three representative values of $\delta_{\rm  sys}$ at 1\%, 5\% and 10 \%, respectively. 
\par The analysis is performed for both hadronic and leptonic observables which depend on the distributions shown in Figures \ref{metXhad} and \ref{metXlep}. Using this definition of $\chi^2$ in \eqref{chidef}, we draw the exclusion contours on the six different two dimensional planes defined by  the anomalous couplings $\left\vert V_{tb}\right\vert\Delta f_1^L$, $ f_1^R$, $f_2^L$ and $ f_2^R$. 68.3\%  and 95\% C.L. Exclusion contours for the hadronic and leptonic channels  are shown in Figures \ref{hadron_bincont68}, \ref{hadron_bincont95} and \ref{lepton_bincont68}, \ref{lepton_bincont95}, respectively. 
For each pair of the couplings, the effect of the overall systematic uncertainty (includes luminosity measurement error etc.) is sketched for three representative values of $\delta{\rm Sys}$ = 1\%, 5\% and 10\%.
\par On examination of the   exclusion contours in both decay modes we find that
\begin{enumerate}
\item[(a)] The sensitivity of measuring all anomalous couplings are affected by the  systematic uncertainty $\delta_{\rm Sys}$.

\item[(b)] The sensitivity of $\left\vert V_{tb}\right\vert\Delta f_1^L$ at 95\% C.L. is of  the  $\sim 5 \times 10^{-3}$ and $\sim 3 \times 10^{-2}$  with systematic error of 1\% and 10 \%, respectively. The order of the sensitivity for other anomalous couplings  varies as $\sim 10^{-2}- 10^{-1}$ at 95 \% C.L. with the $\delta_{\rm Sys}$ varying between .01 to 0.1. 
\end{enumerate}

\subsection{Errors and correlations}
\label{errcorr}
In order to constrain  anomalous $Wtb$ couplings   further  we adopt the  method of optimal observables \cite{Dutta:2008bh,Nachtmann:2004fy} by  using the full information from the distribution of kinematic observables. This technique of estimating the {\it equivalent  maximum likelihood estimator}  has been  used in experimental analysis to compute   the expected efficiencies in extracting the anomalous couplings from the experimental data \cite{Fanourakis:1997bp,Achard:2004ds,Heister:2001qt}.  In this method,  all the anomalous couplings $f_i$, having  different shape profiles from each other can be constrained  simultaneously. 
For a given integrated Luminosity $L$, the statistical errors in the $f_i$ and the correlations of the errors among  anomalous coupling measurement can be obtained from the $\chi^2$ which is a function of all anomalous couplings. Redefining the $\chi^2$ of  equation \eqref{chidef}  in terms of the two anomalous couplings and the covariance matrix $V$  we have
\begin{eqnarray}
  \chi^2(f_i,f_j) = \chi^2_{\rm min} + \sum_{i,j} (f_i -\bar{f_i}) 
   \left[ V^{-1} \right]_{ij} (f_j - \bar{f_j})
  \label{chisq_fifj}
\end{eqnarray}
A total of ten inverse  covariant matrices $V^{-1}$ can be generated
from six and four distinct distributions of kinematic observables in hadronic and leptonic modes, respectively, using the approximation  \eqref{expsm}.   
\begin{onecolumn}
\begin{table}[htb]\footnotesize
\centering
\begin{tabular*}{\textwidth}{c@{\extracolsep{\fill}} cccccc}\hline\hline
%\begin{tabular}{c|c|c|c|c|c|c}\hline \hline
& $A_{\Delta \Phi_{\slashed E_T j_1}}$ & $A_{\Delta \Phi_{\slashed E_T \bar b}}$ & $A_{\Delta \Phi_{\slashed E_T W^-}}$ & $A_{\Delta \Phi_{W^-\bar b}}$ & $A_{\theta_{\bar bj_1}}$ & $A_{\Delta \eta_{\bar \bar b j_1}}$   \\ \hline 
SM+$\sum_i {\rm Bkg}_i$                & .532 $\pm$ .003 & .282 $\pm$ .005 & .503 $\pm$ .004 & .799 $\pm$ .003 &  .023  $\pm$ .001   & -.712 $\pm$ .003\\  
$f_{1}^{R} = +.5$ & .327 $\pm$ .004 & .231 $\pm$ .004 & .564 $\pm$ .004 & .778 $\pm$ .003 &  .0005 $\pm$ .004   & -.806 $\pm$ .003 \\  
$f_{2}^{L} = -.5$ & .528 $\pm$ .004 & .082 $\pm$ .004 & .716 $\pm$ .003 & .748 $\pm$ .003 & -.196  $\pm$ .004   & -.868 $\pm$ .002 \\  
$f_{2}^{L} = +.5$ & .390 $\pm$ .005 & .269 $\pm$ .004 & .585 $\pm$ .004 & .683 $\pm$ .004 &  .106  $\pm$ .005   & -.795 $\pm$ .003\\ 
$f_{2}^{R} = +.5$ & .330 $\pm$ .004 & .363 $\pm$ .004 & .566 $\pm$ .003 & .656 $\pm$ .003 & -.197  $\pm$ .004   & -.823 $\pm$ .002\\ \hline \hline
\end{tabular*}
\caption{\small \em{Asymmetries and its error associated with the kinematic distributions in Figure \ref
{metXhad}  at an integrated  luminosity $L$ = 100 fb$^{-1}$. These asymmetries are computed for a representative value of the anomalous coupling 0.5 along with SM. }}
\label{asym_had}
\end{table}
%\end{onecolumn}
 \begin{table}[htb]\footnotesize
\centering
\begin{tabular*}{\textwidth}{c@{\extracolsep{\fill}} ccccc}\hline\hline
%\begin{tabular}{c|c|c|c|c|c|c}\hline \hline
& $A_{\Delta \Phi_{\slashed E_T l_1}}$ & $A_{\Delta \Phi_{\slashed E_T \bar b}}$ & $A_{\theta_{\bar bl_1}}$ & $A_{\Delta \eta_{\bar b  l_1}}$   \\ \hline 
SM + $\sum_i {\rm Bkg}_i$               & .384 $\pm$ .004  & .710  $\pm$ .003 & .551 $\pm$ .006 & -.765 $\pm$ .007 \\
$f_{1}^{R} = +.5$ & .484 $\pm$ .004  & .702  $\pm$ .003 & .332 $\pm$ .006 & -.821 $\pm$ .003 \\
$f_{2}^{L} = -.5$ & .526 $\pm$ .004  & .620  $\pm$ .003 & .410 $\pm$ .006 & -.831 $\pm$ .002 \\
$f_{2}^{L} = +.5$ & .353 $\pm$ .005  & .812  $\pm$ .003 & .392 $\pm$ .007 & -.850 $\pm$ .003 \\
$f_{2}^{R} = +.5$ & .424 $\pm$ .004  & .684  $\pm$ .003 & .507 $\pm$ .005 & -.809 $\pm$ .003 
\\ \hline \hline
\end{tabular*}
\caption{\small \em{Asymmetries and its error associated with the kinematic distributions in Figure \ref
{metXlep}  at an integrated  luminosity $L$ = 100 fb$^{-1}$. These asymmetries are computed for a representative value of the anomalous coupling 0.5 along with SM and all background processes.  }}
\label{asym_lep}
\end{table}
 \begin{figure*}[htb]
  \centering
  \begin{tabular}{cc}
  \includegraphics[width=0.5\textwidth,clip]{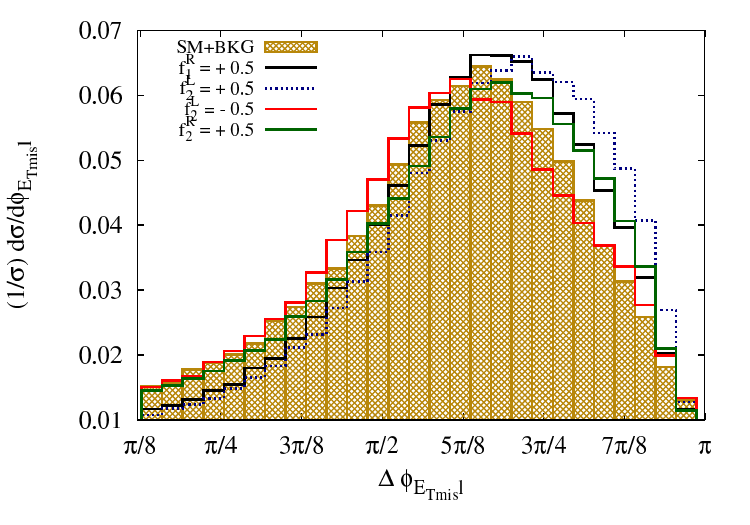} &
  \includegraphics[width=0.5\textwidth,clip]{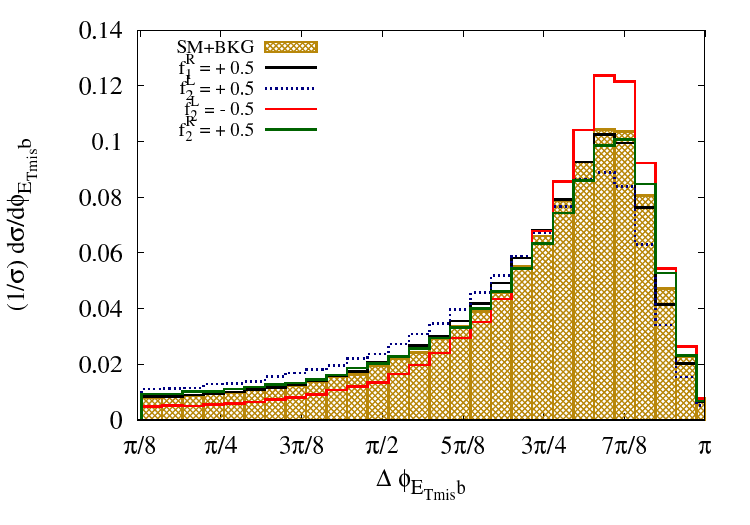} \\
 \includegraphics[width=0.5\textwidth,clip]{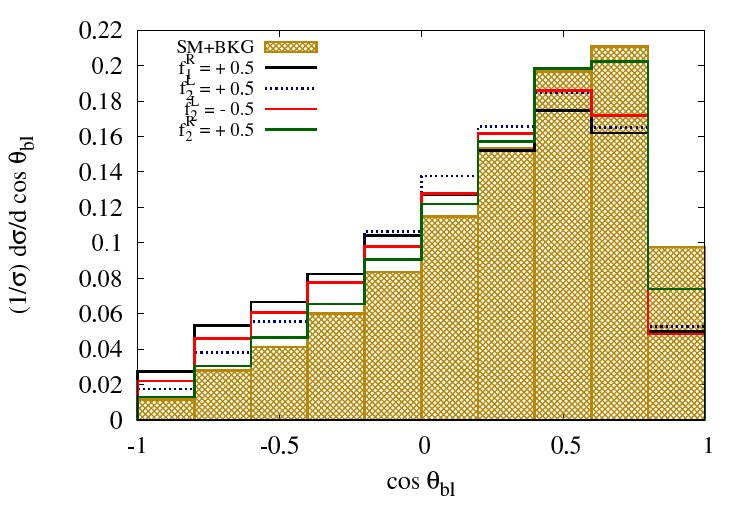} &
 \includegraphics[width=0.5\textwidth,clip]{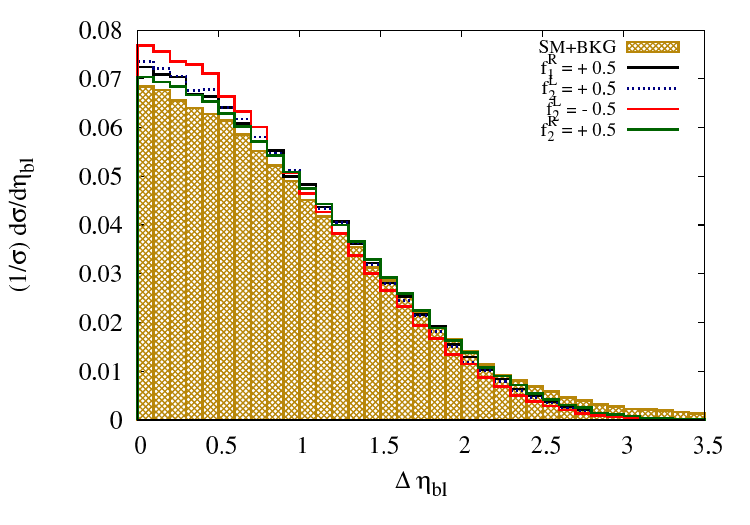}
  \end{tabular}
\caption{\small \em{Normalized distributions of $\Delta\phi_{\slashed E_T\, l_1}$,  $\Delta\phi_{\slashed E_T\, \bar b}$,  $\cos\theta_{\bar b\, l_1}$ and $\Delta\eta_{\bar b\, l_1}$  for leptonic decay mode of $W^-$ corresponding to SM and an anomalous coupling of 0.5. Here $l_1$ is the highest $p_T$ charged lepton. The normalized distributions corresponding to $\left\vert V_{tb}\right\vert \Delta f_1^L=\pm 0.5$ is identical to that of SM. All kinematic observables are measured in lab frame.}}
\label{metXlep}
\end{figure*}
\end{onecolumn}

%The asymmetries induced by the tiny new physics couplings among these two large bins is likely to be seen.
\begin{figure*}[h!]
  \centering
  \begin{tabular}{cc}
  \includegraphics[width=0.5\textwidth,clip]{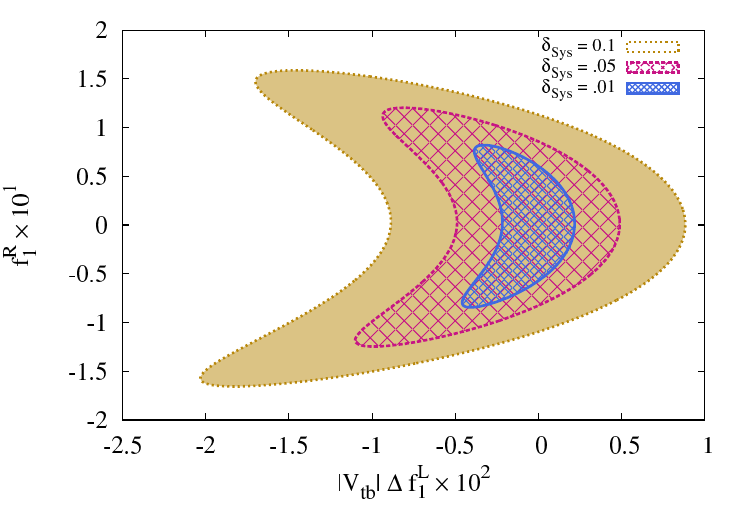} &
  \includegraphics[width=0.5\textwidth,clip]{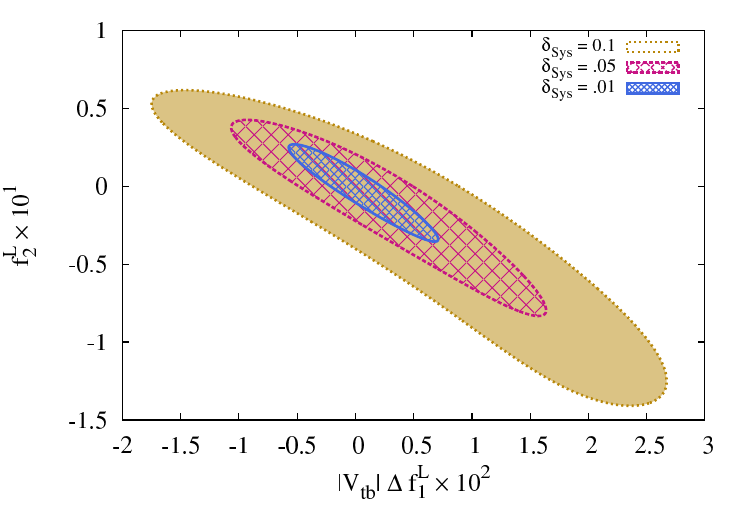} \\
  \includegraphics[width=0.5\textwidth,clip]{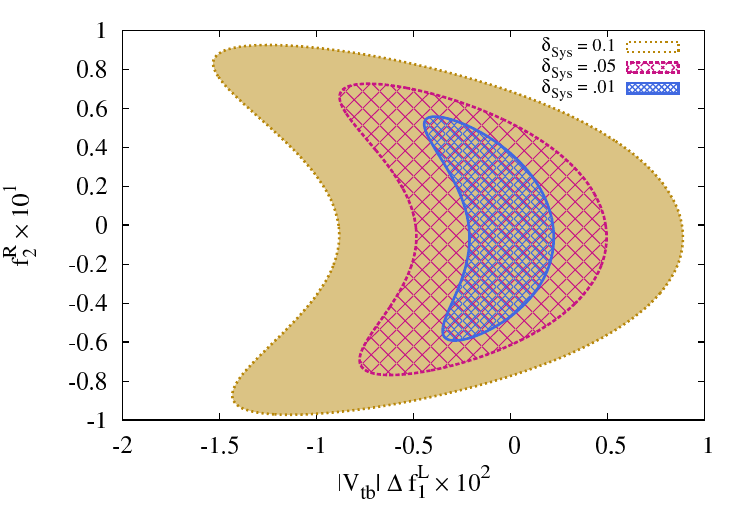} &
  \includegraphics[width=0.5\textwidth,clip]{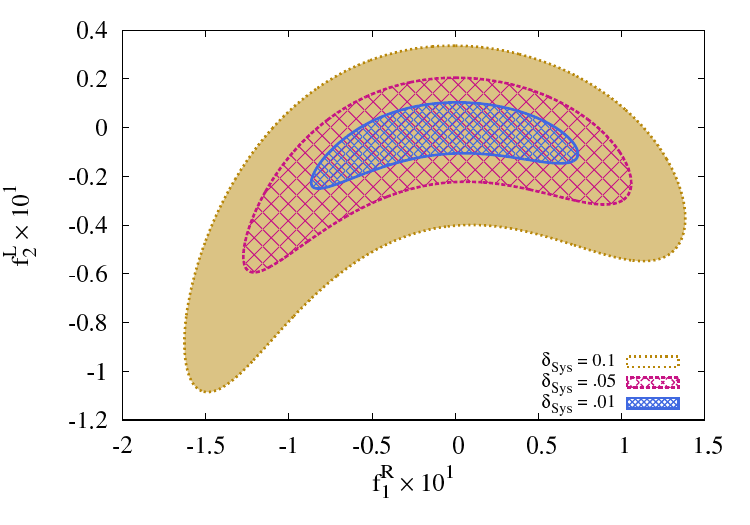} \\
  \includegraphics[width=0.5\textwidth,clip]{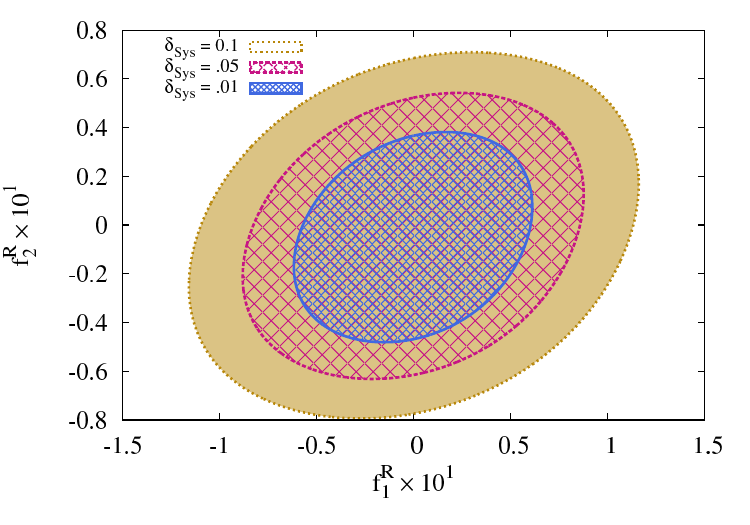} &
  \includegraphics[width=0.5\textwidth,clip]{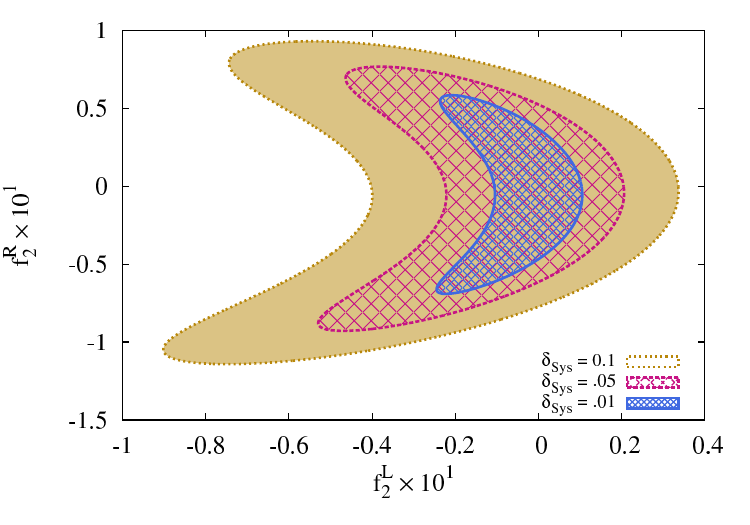} 
  \end{tabular}
\caption{\small \em{68.3 \% C.L. exclusion contours on the plane of $\left\vert V_{tb}\right\vert\Delta f_1^L-f_1^R$, $\left\vert V_{tb}\right\vert\Delta f_1^L-f_2^L$, $\left\vert V_{tb}\right\vert\Delta f_1^L-f_2^R,  f_1^R-f_2^L$, $f_1^R-f_2^R$ and $f_2^L-f_2^R$ and based on combined bin analysis of all  kinematic observables in the hadronic decay mode of $W^-$. A $\chi^2$  analysis is performed by taking into account the deviation from SM and background process with the  systematic error of 1\%, 5\% and 10\%, respectively at an integrated luminosity of $L=100$ fb$^{-1}$.}}
\label{hadron_bincont68}
\end{figure*}
\begin{figure*}[h!]
  \centering
  \begin{tabular}{cc}
  \includegraphics[width=0.5\textwidth,clip]{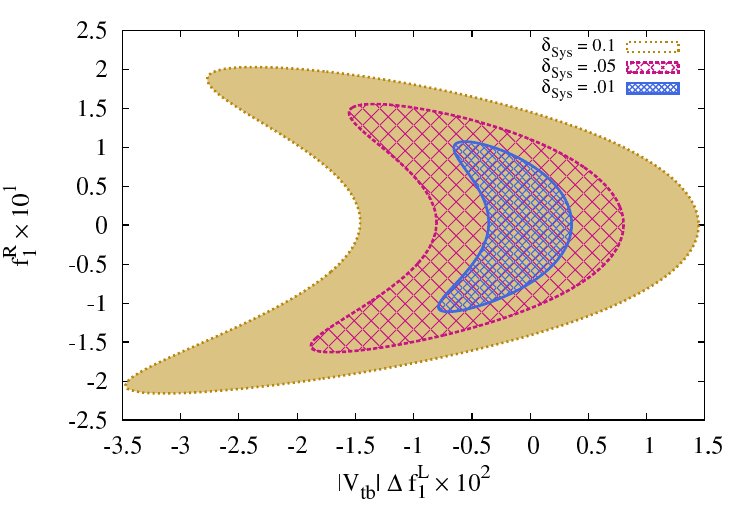} &
  \includegraphics[width=0.5\textwidth,clip]{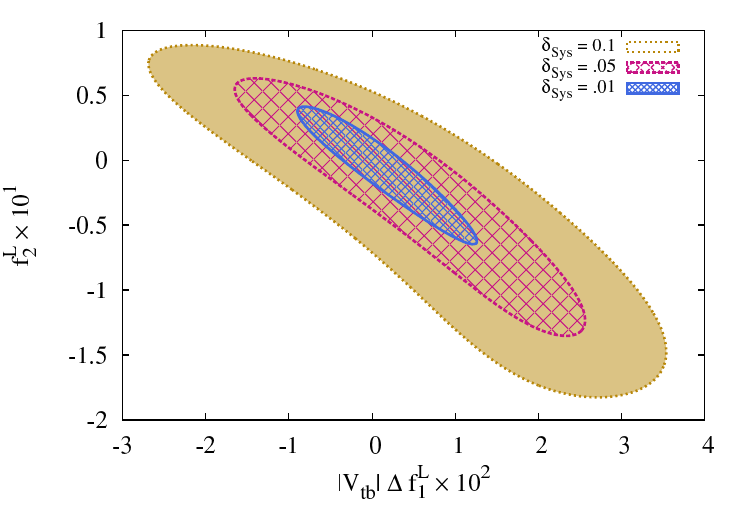} \\
  \includegraphics[width=0.5\textwidth,clip]{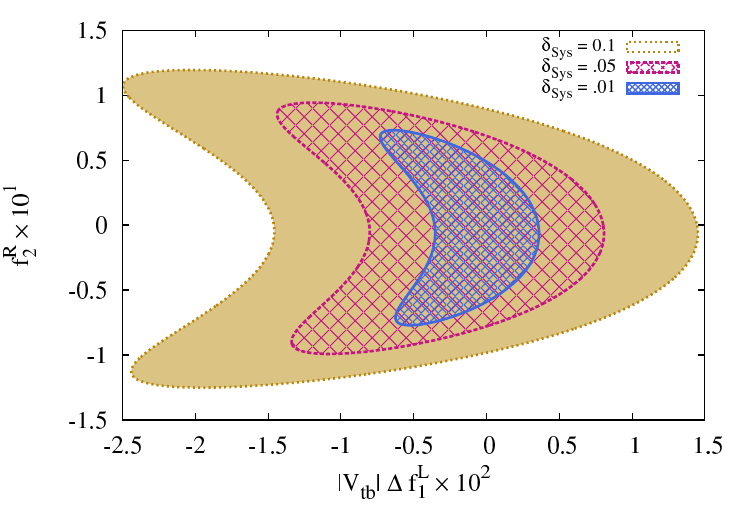} &
  \includegraphics[width=0.5\textwidth,clip]{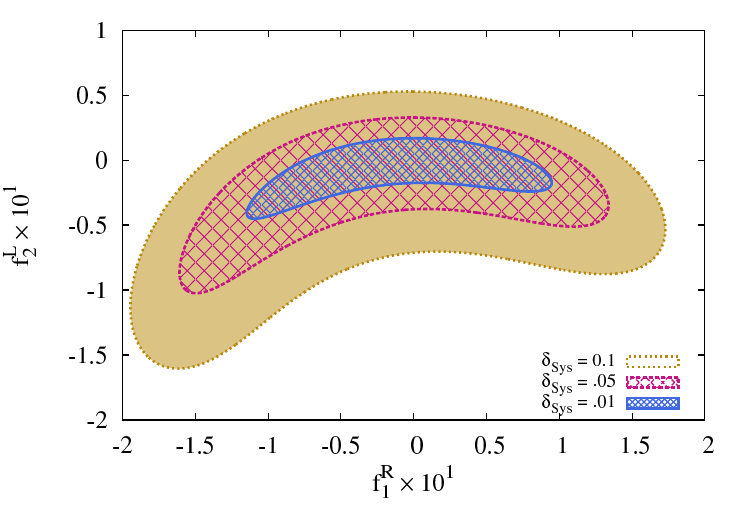} \\
  \includegraphics[width=0.5\textwidth,clip]{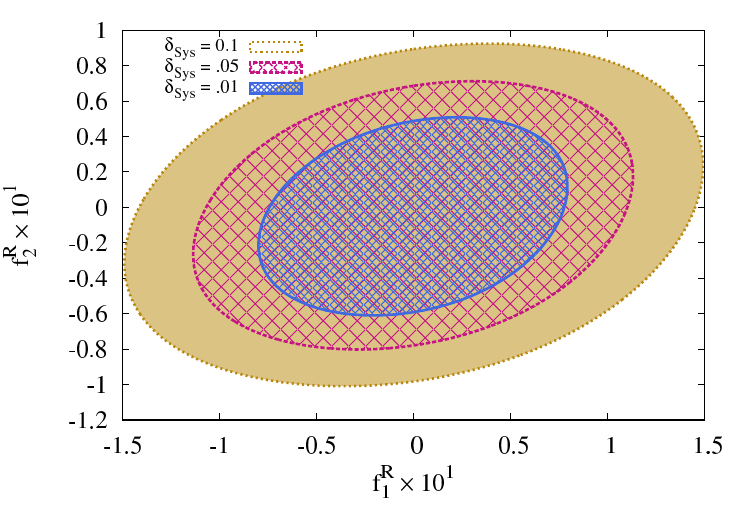} &
  \includegraphics[width=0.5\textwidth,clip]{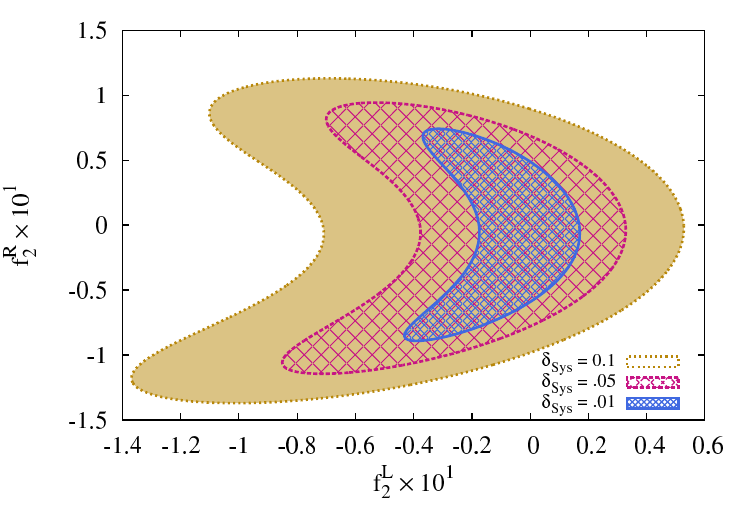} 
  \end{tabular}
\caption{\small \em{95 \% C.L. exclusion contours on the plane of $\left\vert V_{tb}\right\vert\Delta f_1^L-f_1^R$, $\left\vert V_{tb}\right\vert\Delta f_1^L-f_2^L$, $\left\vert V_{tb}\right\vert\Delta f_1^L-f_2^R,  f_1^R-f_2^L$, $f_1^R-f_2^R$ and $f_2^L-f_2^R$ and based on combined bin analysis of all  kinematic observables in the hadronic decay mode of $W^-$. A $\chi^2$  analysis is performed by taking into account the deviation from SM and background process with the  systematic error of 1\%, 5\% and 10\%, respectively at an integrated luminosity of $L=100$ fb$^{-1}$.}}
\label{hadron_bincont95}
\end{figure*}
\begin{figure*}[h!]
  \centering
  \begin{tabular}{cc}
  \includegraphics[width=0.5\textwidth,clip]{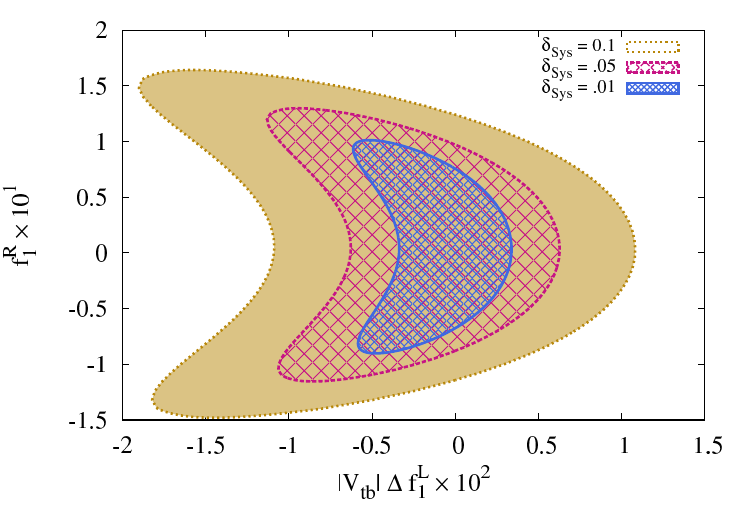} &
  \includegraphics[width=0.5\textwidth,clip]{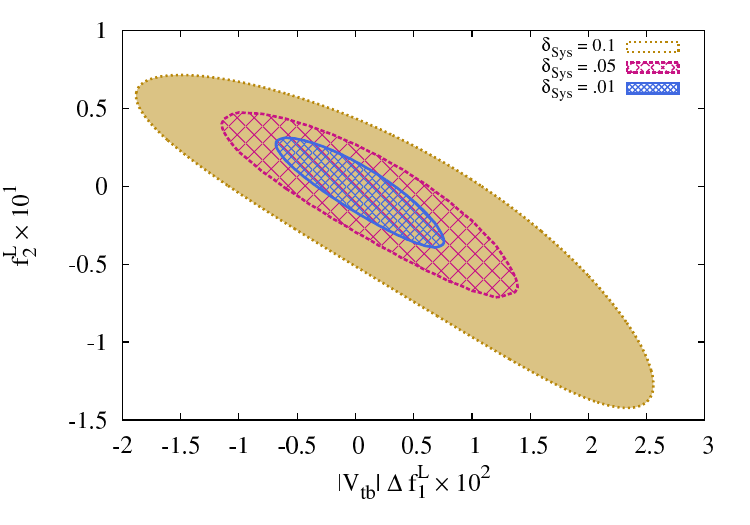} \\
  \includegraphics[width=0.5\textwidth,clip]{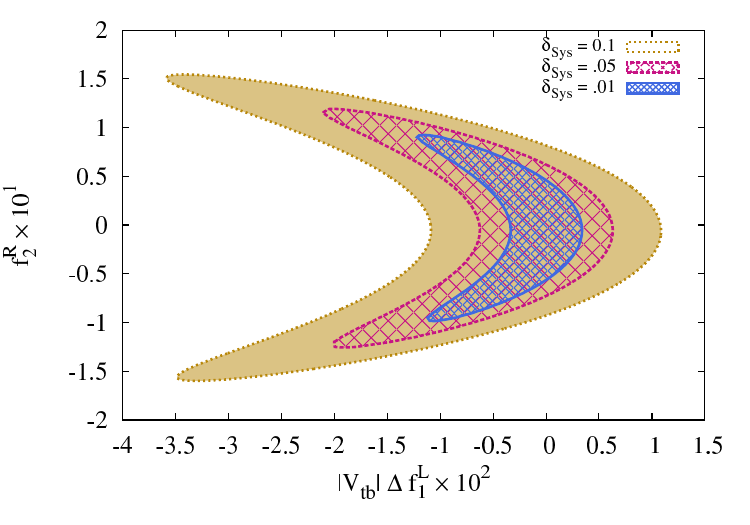} &
  \includegraphics[width=0.5\textwidth,clip]{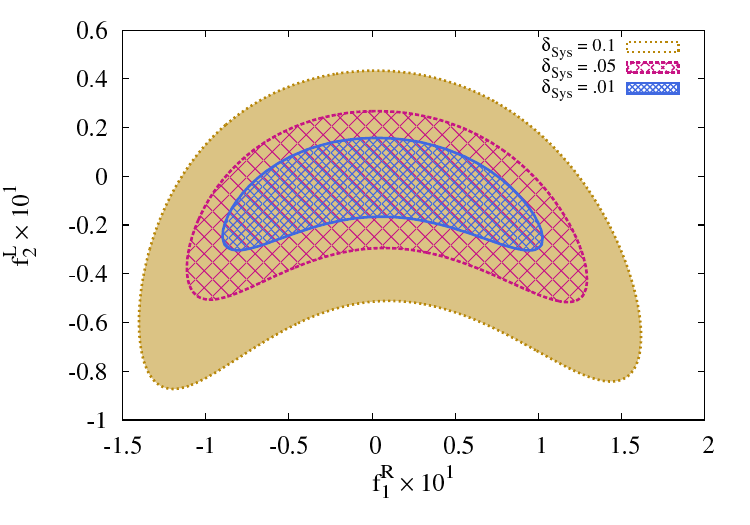} \\
  \includegraphics[width=0.5\textwidth,clip]{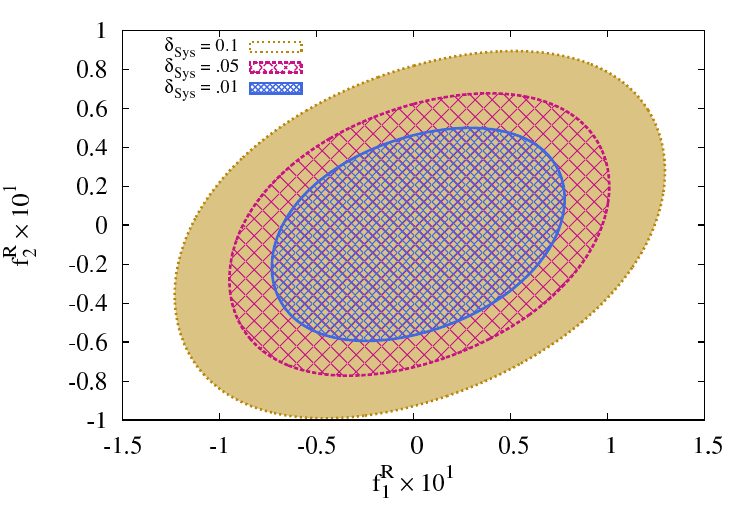} &
  \includegraphics[width=0.5\textwidth,clip]{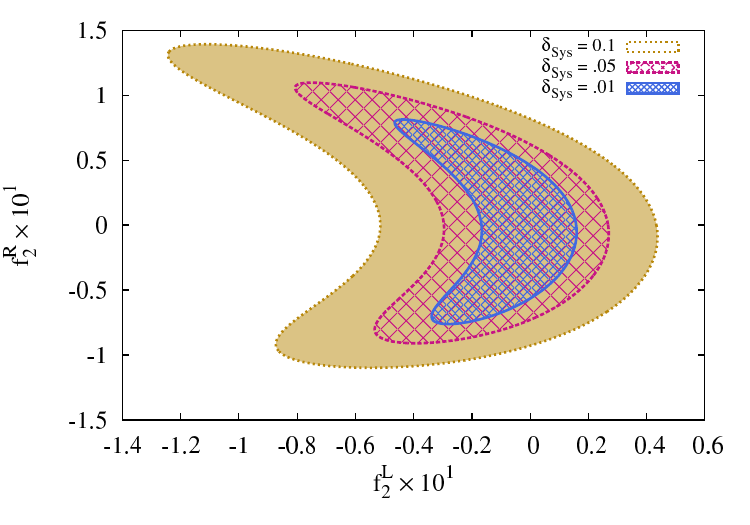} 
  \end{tabular}
\caption{\small \em{68.3 \% C.L. exclusion contours on the plane of $\left\vert V_{tb}\right\vert\Delta f_1^L-f_1^R$, $\left\vert V_{tb}\right\vert\Delta f_1^L-f_2^L$, $\left\vert V_{tb}\right\vert\Delta f_1^L-f_2^R,  f_1^R-f_2^L$, $f_1^R-f_2^R$ and $f_2^L-f_2^R$ and based on combined bin analysis of all  kinematic observables in the leptonic decay mode of $W^-$. A $\chi^2$  analysis is performed by taking into account the deviation from SM and background process with the  systematic error of 1\%, 5\% and 10\%, respectively at an integrated luminosity of $L=100$ fb$^{-1}$.}}
\label{lepton_bincont68}
\end{figure*}
\begin{figure*}[h!]
  \centering
  \begin{tabular}{cc}
  \includegraphics[width=0.5\textwidth,clip]{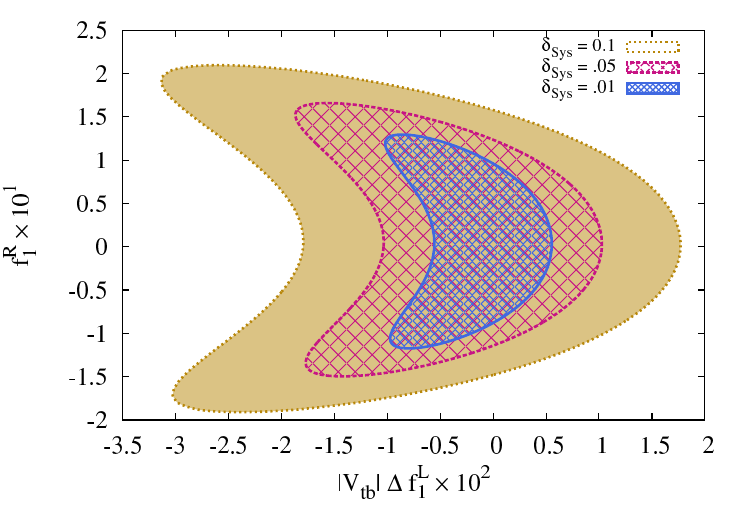} &
  \includegraphics[width=0.5\textwidth,clip]{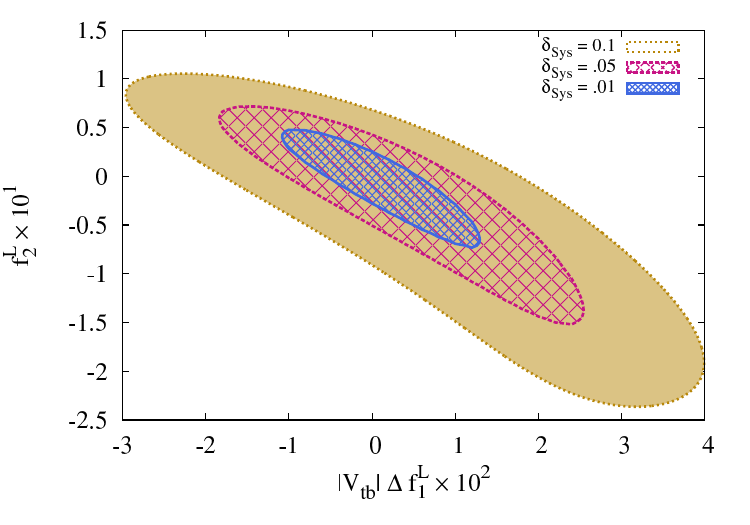} \\
  \includegraphics[width=0.5\textwidth,clip]{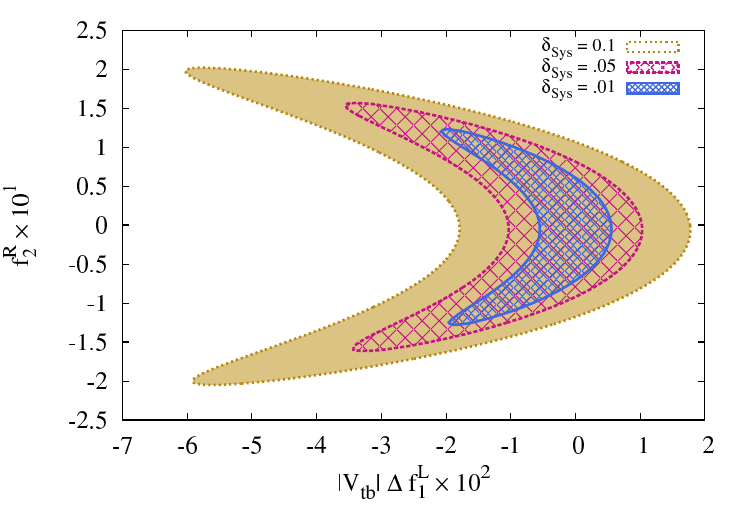} &
  \includegraphics[width=0.5\textwidth,clip]{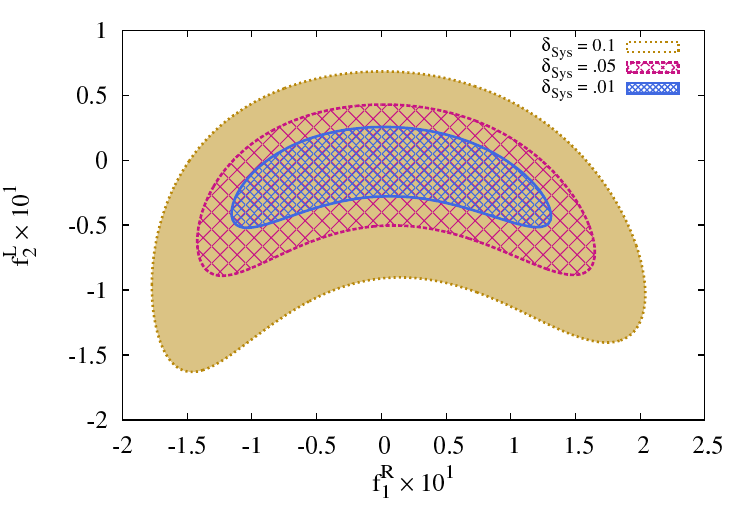} \\
  \includegraphics[width=0.5\textwidth,clip]{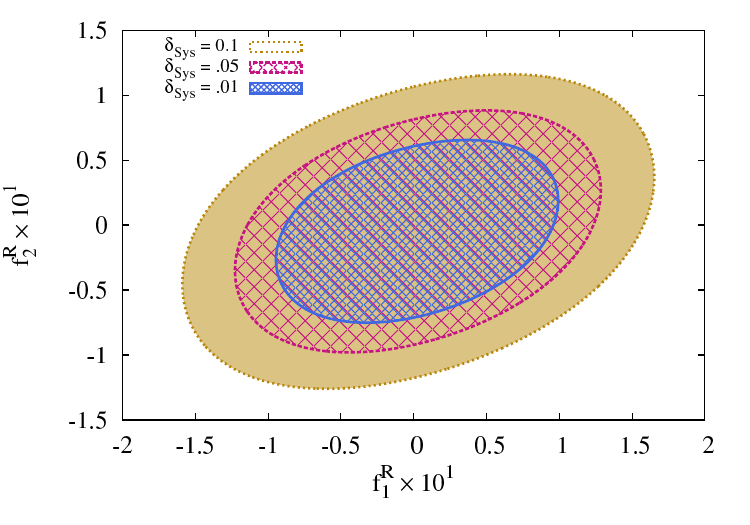} &
  \includegraphics[width=0.5\textwidth,clip]{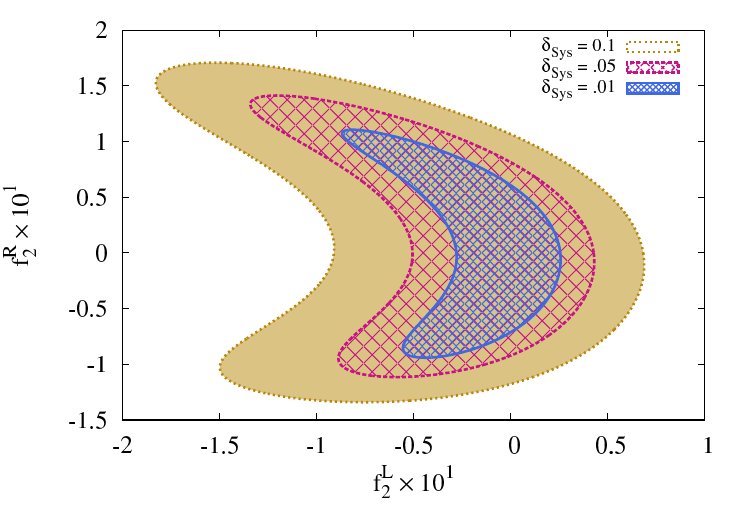} 
  \end{tabular}
\caption{\small \em{95 \% C.L. exclusion contours on the plane of $\left\vert V_{tb}\right\vert\Delta f_1^L-f_1^R$, $\left\vert V_{tb}\right\vert\Delta f_1^L-f_2^L$, $\left\vert V_{tb}\right\vert\Delta f_1^L-f_2^R,  f_1^R-f_2^L$, $f_1^R-f_2^R$ and $f_2^L-f_2^R$ and based on combined bin analysis of all  kinematic observables in the leptonic decay mode of $W^-$. A $\chi^2$  analysis is performed by taking into account the deviation from SM and background process with the  systematic error of 1\%, 5\% and 10\%, respectively at an integrated luminosity of $L=100$ fb$^{-1}$.}}
\label{lepton_bincont95}
\end{figure*}
\begin{twocolumn}
If the SM prediction along with all dominant backgrounds gives a reasonably good
description of the data in most of the phase space region, then the
statistical errors $\Delta f_i$ of $f_i$ and their correlations are determined solely
in terms of the these six covariance matrices $V$ as
\begin{eqnarray}
 f_i - \bar{f_i} = \pm \Delta f_i = \pm \sqrt{V_{ii}}, \hskip 0.5cm
  \rho_{ij} = V_{ij}/\sqrt{V_{ii} V_{jj}}.
\end{eqnarray}
\noindent $\rho_{ij}$ gives correlation coefficient between two distinct anomalous couplings $f_i$ and $f_j$ and gives the absolute error for a given anomalous coupling $f_i=f_j$. $\Delta f_i$ gives the uncertainty with which these couplings will   be measured at the LHeC. 
 $\bar{f_i}$ is the expected mean value in SM, which is zero for all  anomalous couplings $f_i$.
\par  Subsequently, an optimal analysis  with an integrated luminosity $L$ = 100 fb$^{-1}$ is made after combining all kinematic observables in both hadronic and the leptonic modes, respectively. 
\par The inverse of the covariance matrix $V_{ij}^{-1}$ is generated from  one dimensional histogram of each sensitive kinematic  observable and  the corresponding respective correlation matrix is computed. We  combine all inverse covariant matrices to compute the combined $\chi^2$ in the hadronic and leptonic modes separately.

\par The combined $\chi^2$ reads as
\begin{eqnarray}
\chi^2_{\rm comb.}\left(f_i,\,f_j\right)- \sum_{k=1}^n{\chi^2_{\rm min}}_k \nonumber\\
= \sum_k\sum_{i,j} (f_i -\bar{f_i}) 
   \left[ V^{-1} \right]_{ij}^k (f_j - \bar{f_j}) 
\end{eqnarray}
Here $k\equiv $ number of distributions corresponding to the kinematic observables.  $n$ = 6 and 4 for hadronic and leptonic channels, respectively.

We thus provide the accuracy with which  anomalous couplings can be measured from each of these distributions. The correlation matrices and the absolute errors  in each and every couplings in the hadronic and leptonic modes are given below:
\begin{eqnarray}
 \begin{array}{c}
 \left\vert V_{tb}\right\vert \Delta f_1^L = \,\,\pm\,\, 4.5\times 10^{-4} \\
   f_1^R = \,\,\pm\,\, 7.2\times 10^{-4} \\
  f_2^L = \,\,\pm \,\,4.7\times 10^{-4} \\
  f_2^R =\,\,\pm \,\,3.2\times 10^{-4} \\
 \end{array} \label{had_comb._err}
\left(
  \begin{array}{cccc}
   1 \\
   -.07  &  1 \\
   -.04  & -.07  & 1 \\
   -.03  & .006  & -.02 & 1
  \end{array}
 \right);
 \label{had_comb._cormat} \nonumber\\\nonumber\\
({\rm a})\,\,\, { \rm  hadronic\, mode} \hskip 2.5 cm\\
 \begin{array}{c}
  \left\vert V_{tb}\right\vert\Delta f_1^L = \,\,\pm\,\, 4.6\times 10^{-4} \\
   f_1^R = \,\,\pm\,\, 7.2\times 10^{-4} \\
  f_2^L = \,\,\pm \,\,8.3\times 10^{-4} \\
  f_2^R =\,\,\pm \,\,4.3\times 10^{-4} \\
 \end{array} \label{lep_comb._err}
 \left(
  \begin{array}{cccc}
   1 \\
   -.02  &  1 \\
   -.05 & -.06  & 1 \\
  -.01  & .09  & -.07 & 1
  \end{array}
 \right);
 \label{lep_comb0_cormat}\nonumber\\
 \nonumber\\  \hskip 2.5 cm({\rm b})\,\,\, {\rm  leptonic\, mode}\hskip 2.5 cm
\label{cov_lum_err_0}
\end{eqnarray}

\par Up-till now we have considered the hadronic and leptonic modes of single anti-top production at the LHeC to be two different probes for measuring these anomalous couplings. We now combine  observations from both channels  in terms of combined    inverse covariance matrix.  The global errors and correlations from the corresponding global combined covariance matrix  is  then given as 
\begin{eqnarray}
  \begin{array}{c}
 \left\vert V_{tb}\right\vert \Delta f_1^L = \,\,\pm\,\, 3.2\times 10^{-4} \\
   f_1^R = \,\,\pm\,\, 4.6\times 10^{-4} \\
  f_2^L = \,\,\pm \,\,4.2\times 10^{-4} \\
  f_2^R =\,\,\pm \,\,2.6\times 10^{-4} \\
 \end{array} \label{comb.01_err}
 \left(
  \begin{array}{cccc}
   1 \\
   -.05  &  1 \\
   -.04 & -.06  & 1 \\
  -.02  & .03  & -.04 & 1
  \end{array}
 \right);
 \label{comb_had_lep_cormat}
\end{eqnarray} 
It is worthwhile to mention that we have not yet considered any systematic error in the covariance analysis. On comparing the errors given in equations (\ref{cov_lum_err_0}a) and  (\ref{cov_lum_err_0}b) corresponding to hadronic and leptonic modes, respectively and errors for the global combined analysis in equation  \eqref{comb_had_lep_cormat} with those in section \ref{bin analysis}, we find that the  sensitivity of $\left\vert V_{tb}\right\vert \Delta f_1^L$  and others are found to have increased by one  and two orders of magnitude, respectively.

\par We have computed all the errors and their correlations based on  60 \% $b$ tagging efficiency $\epsilon_b$ along with  $10 \% $ and $1\%$   $b$ faking  probability  by charm and light jets respectively. One can however, take these parameters in the $\chi^2$ analysis explicitly rather than  as an overall multiplying factor  in the respective cross-sections.   Since, we consider processes with same final states (same number of $b,\, \bar b$) for the signal as well as the dominant SM top background,  the optimal analysis shows that the sensitivity of the errors in the measurement of these couplings will scale as $1/\sqrt{\epsilon_b}$ for a given luminosity and $\chi^2$. 
\par Assuming the measured luminosity to be the true luminosity, the accuracy with which the anomalous couplings are measured scales as $1/\sqrt{L}$.

\subsubsection{Luminosity Error}
An error in the measurement of luminosity is however, 
is likely to  affect the measurements of some effective couplings. It is thus instructive to study the impact of uncertainty in luminosity measurement on the sensitivity of anomalous $Wtb$ couplings. The true luminosity $L$ can be estimated as
\begin{eqnarray}
L \equiv \beta\bar L, \beta = 1\pm \Delta \beta,
\end{eqnarray}
where $\bar L$ is the measured mean value, and $\Delta \beta$ is its one $\sigma$	
uncertainty. 
 With the inclusion of the luminosity uncertainty the $\chi^2_{\rm comb.}$ definition  given in \eqref{chisq_fifj} is modified to 
\begin{eqnarray}
  \chi^2_{\rm comb.}(f_i,f_j) \rightarrow \chi^2_{\rm comb.}(f_i,f_j,\,\beta ) \equiv  \nonumber\\\sum_{k=1}^m\sum_{i=0}^n\sum_{j=0}^n (f_i -\bar{f_i}) 
   \left[ V^{-1} \right]_{ij}^k (f_j - \bar{f_j}) + \left(\frac{\beta_k -1}{\Delta\beta_k}\right)^2\nonumber\\
  \label{chisq_fifj_lum}
\end{eqnarray}
Here $\left[ V^{-1} \right]_{ij}^k $ is now $ (n+1) \,\times (n+1)$ matrix with $f_0 = \beta -1$. The luminosity
uncertainty $\Delta \beta_k\equiv\Delta \beta$ is same for all kinematic observables at a given collision energy. Here $n\equiv0,1,2,3,4$ corresponding to luminosity factor $\beta$ and four anomalous couplings. $m\,=\,6\,(4)$ corresponds to the number of kinematic observables  for hardonic (leptonic) mode.

\par It is straightforward to integrate out the  $f_0=1-\beta$ 
dependence and obtain the probability distribution of the
parameters $f_1$ to $f_n$ in the presence of the luminosity
uncertainty. $\left\vert V_{tb}\right\vert\Delta f_1^L$ is the only coupling whose weight function  is identical
to the SM distribution at tree level. The other effective couplings get the  SM contribution at the one-loop level and it is thus likely that  the statistical errors  dominate over systematics. Therefore errors coming from the luminosity uncertainty can then be safely neglected for the other three couplings namely $f_1^R$, $f_2^L$ and $f_2^R$.
\par The
impact of the luminosity uncertainty can thus be accounted  algebraically by
using the $\chi^2$ functions written in term of $\Delta f_1^L$.
Redefining our $\chi^2_{comb.}$ function as 
\begin{eqnarray}
\chi^2_{\rm comb.}(f_i,f_j,\,\beta ) =\hskip 5 cm \nonumber\\
 \chi^2_{\rm comb.}\left(\Delta f_1^L\rightarrow 
{\Delta f_1^L}^\prime = \Delta f_1^L + \frac{\beta -1}{2}\right)+\left(\frac{\beta -1}{\Delta\beta}\right)^2\nonumber\\
=\sum_{k=1}^m\sum_{i=0}^n\sum_{j=0}^n f_i^\prime  
   \left[ V^{-1} \right]_{ij}^k f_j^\prime  + \left(\frac{\beta_k -1}{\Delta\beta_k}\right)^2\hskip 2 cm
\label{redefine_lum_chisq}
\end{eqnarray}
\noindent where ${\Delta f_1^L}^\prime= \Delta f_1^L + (\beta-1)/2$ and $f_i^\prime\equiv f_i\, ({\rm for}\, i\ne 1)$
The luminosity uncertainty
in the $\chi^2_{\rm comb.}$ function in equation \eqref{redefine_lum_chisq} can be factored out as
\begin{eqnarray}
\chi^2_{\rm comb.}=\left[\frac{\beta-1}{\Delta \beta^{\rm eff}} +\Delta\beta^{\rm eff}\, R\right]^2+\tilde \chi^2_{\rm comb.},\quad\quad {\rm where}\nonumber \\
\left\{\Delta \beta^{\rm eff}\right\}^{-2}=\frac{1}{\Delta \beta^2}+\frac{1}{4}\,[V^{-1}]_{11};  R=\frac{1}{2}\sum_{a=1}^{4}f_a\, [V^{-1}]_{1a};\nonumber\\
\end{eqnarray}
The new $\tilde\chi^2_{\rm comb.}$ is the reduced combined $\chi^2$ function, which can be re-written as
\begin{eqnarray}
 \tilde\chi^2_{\rm comb.}=\chi^2_{\rm comb.}-\left(\Delta \beta^{\rm eff}\right)^2\, R^2
\label{reduced_chi_sq}
\end{eqnarray}
\par The reduced $\chi^2$ function can now be used to study the constraints
on the effective couplings in the presence of the luminosity
uncertainty. It is worth mentioning that correlations between the
$\left\vert V_{tb}\right\vert\Delta f_1^L$ with other  couplings  are affected due to the presence of the second term in equation \eqref{reduced_chi_sq}.
\par Following the optimal analysis by incorporating the luminosity uncertainty  and the reduced $\tilde\chi^2_{\rm comb.}$, we get $4\times 4$ 
covariance matrix. The  modified correlation matrices based on the combined study of the six and four kinematical distributions from hadronic and leptonic modes, respectively  at an integrated luminosity of $L=100$ fb$^{-1}$ can now be computed for different luminosity uncertainty factor $\beta$.
We give  a spectrum of three correlation matrices corresponding to the three choices  for $\Delta \beta=$ at 1\%, 5\% and 10\%, respectively:
\begin{eqnarray}
 \begin{array}{c}
  \left\vert V_{tb}\right\vert \Delta f_1^L = \,\,\pm\,\, 5.0\times 10^{-3} \\
   f_1^R = \,\,\pm\,\, 4.7\times 10^{-4} \\
  f_2^L = \,\,\pm \,\,4.2\times 10^{-4} \\
  f_2^R =\,\,\pm \,\,2.6\times 10^{-4} \\
 \end{array} \label{comb.01_lumin_err}
 \left(
  \begin{array}{cccc}
    1 \\
    -.003  &  1 \\
    -.003  &  -.068  & 1 \\
    -.002  & .032  & -.041 & 1
  \end{array}
 \right);\label{comb.01_lumin_cormat}\nonumber\\
({\rm a})\,\,\, \Delta\beta = 1\% \hskip 2.5 cm\\
 \begin{array}{c}
  \left\vert V_{tb}\right\vert\Delta f_1^L = \,\,\pm\,\, 2.5\times 10^{-2} \\
   f_1^R = \,\,\pm\,\, 4.6\times 10^{-4} \\
  f_2^L = \,\,\pm \,\,4.2\times 10^{-4} \\
  f_2^R =\,\,\pm \,\,2.6\times 10^{-4} \\
 \end{array} \label{comb.05_lumin_err}
 \left(
  \begin{array}{cccc}
   1 \\
   0  &  1 \\
   0  & -.068  & 1 \\
   0  & .032  & -.041 & 1
  \end{array}
 \right);
 \label{comb.05_lumin_cormat}\nonumber\\
\nonumber\\
({\rm b})\,\,\, \Delta\beta = 5\% \hskip 2.5 cm
\label{cov_lum_err_.01_.05}
\end{eqnarray}
\begin{eqnarray} 
 \begin{array}{c}
  \left\vert V_{tb}\right\vert\Delta f_1^L = \,\,\pm\,\, 5.0\times 10^{-2} \\
   f_1^R = \,\,\pm\,\, 4.6\times 10^{-4} \\
  f_2^L = \,\,\pm \,\,4.2\times 10^{-4} \\
  f_2^R =\,\,\pm \,\,2.6\times 10^{-4} \\
 \end{array} \label{comb.1_lumin_err}
 \left(
  \begin{array}{cccc}
   1 \\
   0  &  1 \\
   0  & -.068  & 1 \\
   0  & .032  & -.041 & 1
  \end{array}
 \right);
 \label{comb.1_lumin_cormat}\nonumber\\
\nonumber\\
({\rm c})\,\,\, \Delta\beta = 10\% \hskip 1.75 cm
\label{global_corr_1_5_10}
\end{eqnarray}
It is observed from equations \eqref{comb.01_lumin_cormat}, \eqref{comb.05_lumin_cormat}  and \eqref{comb.1_lumin_cormat}   that the sensitivity of all couplings except   $\left\vert V_{tb}\right\vert\Delta f_1^L$ remain same as before given in \eqref{comb_had_lep_cormat}.  The sensitivity of $\left\vert V_{tb}\right\vert\Delta f_1^L$ which has same weight function as SM is however, reduced by one order of magnitude $\sim 10^{-3}$ corresponding to luminosity uncertainty 1\%.  The error in $\left\vert V_{tb}\right\vert\Delta f_1^L$ is now comparable to that  obtained in the bin analysis  with 1\% systematic error.  Following the same suite of bin analysis the sensitivity further worsens by an order of magnitude $\sim 10^{-2} $ with increased luminosity uncertainty at  5\% -10\% uncertainty. 
\par On assumption that the statistical error might dominate over the systematics in the determination of all other couplings, we observe that they are not affected due to the varying $\Delta\beta$  as    mentioned in the  definition of  $\chi^2_{\rm comb}$. This is in sharp contrast to that observed in bin analysis where all couplings are affected by the systematic uncertainty. 
\par The correlations of $f_1^R$, $f_2^L$ and $f_2^R$ with $\left\vert V_{tb}\right\vert\Delta f_1^L$ are drastically reduced for $\Delta \beta$ = .01 and finally becomes vanishingly small for
$\Delta \beta$ = .05 and $\Delta \beta$ = 0.10. However, the correlations among   $f_1^R$, $f_2^L$ and $f_2^R$ remain same as given in equation \eqref{comb_had_lep_cormat}.
\begin{figure}
\centering
  \includegraphics[width=0.5\textwidth,clip]{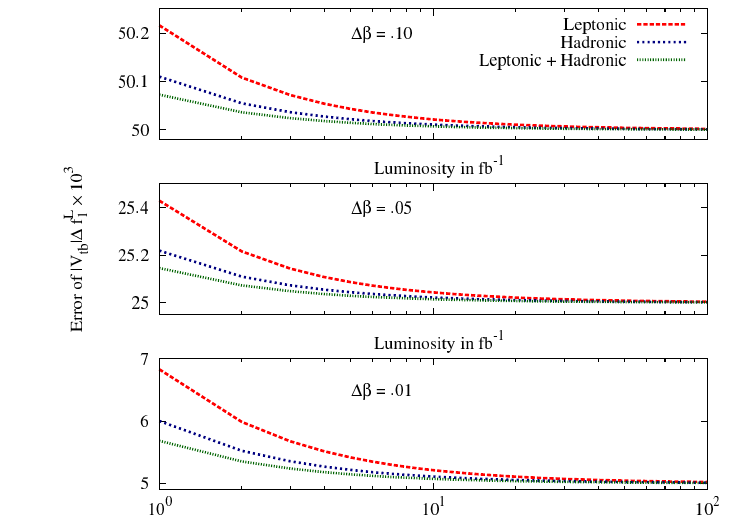} 
  \caption{\small \em{ Variation of the total error of $\left\vert V_{tb}\right\vert\Delta f_1^L$  with the luminosity for a given luminosity uncertainty of 1\%, 5\% and 10\%, corresponding to the kinematical  distributions from hadronic, leptonic and combined decay modes of $W^-$, respectively.}}
\label{lumn:fig}
\end{figure} 
\par As an illustration, we  study  the variation in total error measurement of $\left\vert V_{tb}\right\vert\Delta f_1^L$ based on this optimal analysis with a fixed  luminosity uncertainty $\Delta \beta$.
 In Figure  \ref{lumn:fig},  the variation of the total error in the estimation of $\left\vert V_{tb}\right\vert\Delta f_1^L$, with the luminosity for a given $\Delta\beta$ is shown. 
Thus the error in the anomalous coupling  not only depends  on the high magnitude of the luminosity, but also on its measured value. 
\section{Observations and Discussion}
\label{summary}
An attempt has been made to study and investigate  the sensitivity of the measurement of anomalous $Wtb$ couplings associated with the left or right  vector and tensor chiral currents. The LHeC being comparatively clean with respect to $pp$ and $p\bar p$  colliders, provides an excellent environment to study the electroweak production of single anti-top. We analyse the effect of anomalous couplings in the $Wtb$ vertex by examining its  one dimensional  distributions.
\subsection{Observations} We    summarize our results as follows :
\begin{enumerate}
\item[(i)]  It is found that  asymmetries of kinematic variables constructed from respective one dimensional distributions   can  discriminate the effect of non-SM contribution through new  vector and tensor chiral currents except for $\left\vert V_{tb}\right\vert\Delta f_1^L$, as shown in Tables \ref{asym_had} and \ref{asym_lep},  for  anomalous couplings are of the order of $\sim 10^{-1}$. The asymmetry study suggests that the distribution of the cosine of angle between  the tagged $\bar b$ quark and the highest $p_T$ jet $j_1$ in the hadronic decay mode of $W^-$ to be the most sensitive observable. 
\item[(ii)] We have conducted a $\chi^2$ analysis based on the differential events of  all kinematic variables for hadronic and leptonic modes in SM and background channels.    
This gives the exclusion   contours  on the anomalous couplings plane.  Contours at 68 \% and 95 \% are provided for both hadronic and leptonic decay modes of $W^-$ in Figures \ref{hadron_bincont68} and \ref{lepton_bincont68} and   in Figures \ref{hadron_bincont95} and \ref{lepton_bincont95} respectively. 
\par The sensitivity of $\left\vert V_{tb}\right\vert\Delta f_1^L$ at 95\% C.L. is found to be of the order of  $\sim  10^{-3} - 10^{-2}$  with the corresponding variation of 1\% - 10\% in the systematic error (which includes the luminosity error). The order of the sensitivity for other anomalous couplings  varies between $\sim 10^{-2}- 10^{-1}$ at 95 \% C.L. 
\par We find that the sensitivity of the anomalous couplings can be increased with the increase in the luminosity as the couplings scales as $ ~1/\sqrt{L}$ for a given $\chi^2$. Thus for 1 ab$^{-1}$ the sensitivity of all $f_i$'s are going to be roughly improved  by a factor of $\approx 0.31$. Similarly, for a given  integrated true luminosity and $\chi^2$, an $n$ fold increase in $b$-tagging efficiency  would increase the sensitivity of the anomalous couplings by $1/\sqrt{n}$.  

\item[(iii)] Adopting   the technique of the optimal observable, we obtained the global combined error sensitivity of all couplings is of the order of $\sim 10^{-4}$ in the absence of any systematic uncertainty. 
\item[(iv)]   Lastly, we have  extended our optimal analysis to include the luminosity uncertainty factor in addition to four anomalous couplings. The increasing luminosity error  reduces the sensitivity of $\left\vert V_{tb}\right\vert \Delta f_1^L$ and its correlation with other couplings. On combining the results from both hadronic and leptonic modes  and errors with luminosity uncertainty at 1\% we find that the error sensitivity of $\left\vert V_{tb}\right\vert \Delta f_1^L\sim 10^{-3}$ becomes comparable to that observed in the bin analysis. The sensitivity is further reduced to $~10^{-2}$ for luminosity uncertainty greater than 5\%. However, sensitivity of all other couplings are unchanged at $\sim 10^{-4}$. 
\end{enumerate}
\subsection{Comparison and Analysis}
We compare our results with those  quoted in the joint report {\it TOPLHCNOTE}  \cite{atlasnote2013-33}, based on the  recent experimental data at $\sqrt{s}=7$ TeV and integrated  luminosity of 35 pb$^{-1}$ to 2.2 fb$^{-1}$. They found the sensitivity of the $Re(f_2^R) = −0.10 \pm 0.06 (stat.)+ {}^{+0.07}_{-0.08} (syst.)$. Performing the analysis for the LHeC with $E_p=7$ TeV, $E_e=60$ GeV and integrated luminosity of 100 fb$^{-1}$, we find the upper limit on  anomalous coupling $\left\vert f_2^R\right\vert \simeq $ 0.011  and  0.01  for  the hadronic and leptonic modes, respectively.
\par Alternatively, one  constrains the  $C_{tW}/\Lambda^2$, a coefficient of dimension six operator $O_{tW}=\left(\bar q\,\sigma^{\mu\nu}\,\tau^I\,t\right)\,\,\tilde \phi W^{I}_{\mu\nu}$ that contribute to the  $Wtb$ anomalous coupling. By translating the upper bound on the $f_2^R$ on the upper limit of the coefficient corresponding to the dimension six operator, we find  $\left\vert C_{tW}/\Lambda^2\right\vert \leq 0.13$ TeV$^{-2}$. 
\par However, the limit from  low energy electroweak precision data on the above operator is much stronger  $\sim \left\vert C_{tW}/\Lambda^2\right\vert \leq 0.4\pm 1.2$ TeV$^{-2}$ \cite{Zhang:2012cd} than the present LHC bound and till date it provides the benchmark upper limit on the coefficient for this operator. Electro-weak precision data also constrains the other coefficient $C_{bW}/\Lambda^2\leq 11\pm 13$ associated with  dimension six operator $O_{bW}=\left(\bar q\,\sigma^{\mu\nu}\,\tau^I\,b\right)\,\, \phi W^{I}_{\mu\nu}$. Translating the upper bound from the coefficient $f_2^L$, we find that the proposed LHeC will improve the bound to a level of $10^{-2}$ as evident from equation \eqref{global_corr_1_5_10}.  
\par Recently, a detailed study on the top anomalous couplings for LHC at 14 TeV with 10 fb$^{-1}$ data \cite{Bach:2012fb} is done. They have computed the effect of the  anomalous couplings at both production and decay vertices into the full  $t$ channel matrix element of the single top quark production and illustrated  that  one sigma contours on the plane  of the  anomalous couplings lie within  order of magnitude $\sim 10^{-1}$.  Therefore, the accuracy with which these couplings are measured at LHC can then be improved upon in the proposed LHeC  as shown in Figures \ref{hadron_bincont68}, \ref{hadron_bincont95}, \ref{lepton_bincont68}, \ref{lepton_bincont95} from the bin analysis.
\par At present the  stringent upper bound on the magnitude of the anomalous couplings  exist from the low energy $B$ physics experiments as mentioned in the introduction and given in references \cite{Larios:1999au,Burdman:1999fw,Barberio:2007cr, Grzadkowski:2008mf, Drobnak:2011aa}. On comparing with these limits we find that the LHeC might be able to measure these anomalous couplings at the same level of accuracy or even can do better with a high  luminosity facility having  luminosity uncertainty $\le $ 1-2\%.
\par The single top quark production process at ILC is realized through $e^++e^-\to t+\bar b +e^-+\bar\nu_{e^-}$ which is sub-dominant in comparison to the top pair production process $e^++e^-\to t+\bar t$. Therefore, the sensitivity analysis of $Wtb$ anomalous couplings at ILC have been performed in the top pair production processes followed by their decay in hadronic, semi-leptonic and leptonic decay modes \cite{AguilarSaavedra:2012xe,Boos:1999ca,Batra:2006iq}. Boos {\it et. al} \cite{Boos:1999ca} have  simulated the observables forward-backward asymmetry, spin-spin  asymmetry of the top/anti-top decay products and  the asymmetry of the lepton energy spectrum for their analysis of $Wtb$ couplings and found that  $2\sigma$ exclusion contours predict that no distinction can be made with SM if $f_2^L \in [-0.025, 0]$ and $f_2^R\in [-0.20, 0.20]$. Another, study by   Batra and Tait  shows the sensitivity of the anomalous couplings are of the order of $\sim 3\%$ with 100 fb$^{-1}$ data \cite{Batra:2006iq}.    Errors in the measurement at ILC can however,  drastically reduced by improving the top/ anti-top reconstruction tools like $b$-tagging efficiency,  the vertex charge determination and the top identification, though ILC is better suited top explore the  sensitivity of the dimension six operators associated with flavour conserving and flavor changing neutral currents \cite{Baer:2013cma,Amjad:2015mma}.  

\par Our analysis  shows that we can probe the $Wtb$ vertex at the LHeC to a very high  accuracy and can obtain much more stringent upper limits on  anomalous couplings, in comparison to  existing limits from the LHC, electroweak physics and  $B$ meson decays.  The arXiv version of this study  has been discussed in the High Energy Particle Physics Workshop 2015 \cite{Kumar:2015jna}, LHeC Workshop 2015 \cite{LHeC2015} and  in the DIS 2015 wokshop \cite{DIS2015}. We hope that our report will be useful in studying the physics potential of the LHeC project.

\begin{acknowledgements}
% Standard acknowledgements start here
%----------------------------------------------
SD, AG and  MK would like to acknowledge the partial support from DST, India under grant SR/S2/HEP-12/2006.  AG would like to acknowledge  CSIR (ES) award  for the partial financial support.
\end{acknowledgements}
%----------------------------------------------

\bibliographystyle{spphys}    

\end{twocolumn}
\end{document}